\documentclass[trackchanges,twocolumn]{aastex701}
\usepackage{amsmath}

\begin{document}

\title{A Universal Dance of Galactic Disks: Ubiquitous Precession and Its Implications}

\author[orcid=0009-0000-9614-9237,gname=Yuan, sname=Wang]{Yuan Wang}
\email{ywang123@mail.ustc.edu.cn}  
\altaffiliation{These authors contributed equally to this work.}
\affiliation{Department of Astronomy, University of Science and Technology of China, Hefei 230026, Anhui, China}
\affiliation{School of Astronomy and Space Science, University of Science and Technology of China, Hefei 230026, Anhui, China}

\author[orcid=0009-0006-0435-9469,gname=Xiong, sname=Luo]{Xiong Luo} 
\email{luoxiong@mail.ustc.edu.cn} 
\altaffiliation{These authors contributed equally to this work.}
\affiliation{Department of Astronomy, University of Science and Technology of China, Hefei 230026, Anhui, China}
\affiliation{School of Astronomy and Space Science, University of Science and Technology of China, Hefei 230026, Anhui, China} 

\author[orcid=0000-0002-4911-6990,gname=Huiyuan, sname=Wang]{Huiyuan Wang} 
\affiliation{Department of Astronomy, University of Science and Technology of China, Hefei 230026, Anhui, China}
\affiliation{School of Astronomy and Space Science, University of Science and Technology of China, Hefei 230026, Anhui, China}
\email[show]{whywang@ustc.edu.cn}

\author[orcid=0000-0003-1588-9394,gname=Enci, sname=Wang]{Enci Wang} 
\affiliation{Department of Astronomy, University of Science and Technology of China, Hefei 230026, Anhui, China}
\affiliation{School of Astronomy and Space Science, University of Science and Technology of China, Hefei 230026, Anhui, China}
\email[show]{ecwang16@ustc.edu.cn}

\author[orcid=0000-0002-4326-3543,gname=Hao, sname=Li]{Hao Li} 
\affiliation{Department of Astronomy, University of Science and Technology of China, Hefei 230026, Anhui, China}
\affiliation{School of Astronomy and Space Science, University of Science and Technology of China, Hefei 230026, Anhui, China}
\email{lh123@mail.ustc.edu.cn}

\author[orcid=0000-0003-3816-7028,gname=Federico, sname=Marinacci]{Federico Marinacci} 
\affiliation{Departimento di Fisica e Astronomia ``Augusto Righi'', Universit\`a di Bologna, via Piero Gobetti 93/2, Bologna I-40129, Italy}
\affiliation{INAF, Osservatorio di Astrofisica e Scienza dello Spazio di Bologna, via Piero Gobetti 93/3, Bologna I-40129, Italy}
\email{federico.marinacci2@unibo.it}

\author[orcid=0000-0002-6196-823X,gname=Xuejian, sname=Shen]{Xuejian Shen} 
\affiliation{Department of Physics and Kavli Institute for Astrophysics and Space Research, Massachusetts Institute of Technology, Cambridge, MA 02139, USA}
\email{xuejian@mit.edu}

\author[orcid=0000-0001-8593-7692,gname=Vogelsberger, sname=Mark]{Mark Vogelsberger} 
\affiliation{Department of Physics and Kavli Institute for Astrophysics and Space Research, Massachusetts Institute of Technology, Cambridge, MA 02139, USA}
\affiliation{Fachbereich Physik, Philipps-Universität Marburg, D-35032 Marburg, Germany}
\email{mvogelsb@mit.edu}


\begin{abstract}

Precession is a very common phenomenon for small-scale astronomical objects. However,
the precession of galactic disks, occurring on a scale larger than kilo-parsec, has been insufficiently studied in past works. 
Quantifying this precession in observations remains challenging due to the lack of high-resolution dynamical data. Cosmological simulations, where gravitational interactions are self-consistently modeled, offer a unique avenue for investigating disk precession. Leveraging the IllustrisTNG simulations, we trace the evolution of spin orientation in Milky Way-like galaxies over cosmic time. We find that disk precession is ubiquitous in galaxies and significantly affects galaxy evolution. The precession is driven by the external tidal torque originating from the anisotropic matter distribution within $30\ \mathrm{kpc}$, and is violent at $\mathrm{z} > 1$ and becomes gentler but significant at $\mathrm{z} \sim 0$, when the disks are considered dynamically settled. Disk precession can induce significant cold gas warp, which is often observed in the Milky Way and nearby galaxies.
We predict that the Milky Way is precessing at a rate of $\simeq3-10$ degrees per billion years at current epoch based on its observed warp.
Violent precession can heat the orbits of stars, which may eventually produce prolate elliptical galaxies. The tidal torque from central galaxies can cause the precession of nearby satellite galaxies and causes their disks to point towards the centrals, which explains the observational radial alignment. We also find that the precession of accreted cold gas stream, regulated by the galaxies' torque, is crucial for the evolution of disk galaxies.

\end{abstract}


\section{Introduction} \label{sec:intro}

Galactic dynamics is one of the central fields in modern astronomy, providing a crucial understanding of the intricate interaction between dark matter and baryonic matter. Exploring the origin and transport of angular momentum is of fundamental importance \citep{Peebles69, White84, Bell72, Athan03, 2023Romeo}, since these processes play an important role in driving the structural and morphological evolution of galaxies over the course of cosmic history.


Precession, an important process of angular momentum evolution, is an almost ubiquitous phenomenon in small-scale astrophysical systems, spanning from the axial wobble of planets to the intricate behavior of accretion disks around compact objects. In contrast, large-scale precession of stellar disks over several kiloparsecs remains poorly investigated.
Whereas orbital rotation can be readily inferred from Doppler measurements, the precession of a galaxy’s angular momentum vector unfolds over gigayear timescales, making such slow, global motions extremely challenging to quantify observationally. Because any observation provides only a single “snapshot” of an inherently time-dependent dynamical process, it is likewise difficult to clearly separate a steady-state precession from other dynamical configurations, such as those induced by recent merger events.


Although direct observational evidence for galactic disk precession is still scarce, a growing body of indirect astronomical signatures increasingly points to its ubiquity. For instance, \cite{Welker2014} found galaxies embedded in anisotropic environments can experience substantial angular-momentum “flips”, our Milky Way and nearby systems frequently display pronounced disk warps \citep{ han2023NatAs_tilted, Lilleengen2023, huang2024NatAs_slightly}, and satellite galaxies located near their hosts often have disks oriented towards their centrals \citep{pereira2005Radial, faltenbacher2007Three}. While these phenomena have typically been studied in isolation, they may instead constitute different expressions of a common underlying process, galactic precession, as demonstrated in this work. Consequently, adopting a precession-based framework provides a unified dynamical interpretation that connects and reconciles these otherwise disparate observational manifestations within anisotropic environments.


Galactic disk precession arises fundamentally from the non-spherical nature of the dark matter structures that host galaxies. Within the $\Lambda$CDM paradigm, dark matter halos are typically triaxial \citep{Jing2002, Allgood2006, WangH2011}, producing significant gravitational torques on the embedded disks \citep{Zjupa_tidal_2020}. When the disk’s angular momentum is substantially misaligned with the principal axes of its halo, these torques drive a reorientation of the disk plane \citep{2022Dillamore, Zhu2026}. In addition, the hierarchical assembly of galaxies adds further complications: the continual accretion of satellite systems and the influence of massive neighbors such as the Large Magellanic Cloud in the case of the Milky Way introduce external gravitational disturbances that can accelerate or modulate this tilting process \citep{Dodge2023, Vasiliev2023}. Investigating how these various mechanisms jointly determine a galaxy’s overall orientation is crucial for understanding the dynamical history of disk galaxies.

High-resolution cosmological simulations offer a unique and powerful avenue for investigating the physical origins and implications of disk precession. Unlike observations, these simulations allow researchers to track the full trajectory of a galaxy’s angular momentum vector across cosmic time \citep{Zjupa2017, Rodriguez2022}, self-consistently modeling the complex interplay between gravitational interactions and gas physics within a realistic hierarchical environment. These simulations provide a statistical laboratory where one can isolate the influence of various factors, such as anisotropic mass distribution, star formation and baryonic processes, and merger events. It is crucial for interpreting the sparse kinematic data currently available from Gaia and other spectroscopic surveys. Motivated by these needs, we present a systematic investigation of the galactic disk precession utilizing the high-resolution IllustrisTNG simulations \citep{Nelson2019TNG50, Pillepich2019TNG50}. In this work, we quantify the precession rates of Milky Way-like galaxies and analyze the underlying torques from dark matter haloes and satellite interactions, aiming to provide a theoretical framework for future kinematic observations.


This paper is organized as follows: In Section~\ref{sec:method}, we describe the simulation and our sample selection; Section~\ref{sec:origin} analyzes the ubiquity and origin of disk precession; Section~\ref{sec:warp} discusses the induction of galactic warps; Section~\ref{sec:satellite} examines satellite alignment; Section~\ref{sec:morphology} analyzes the impact of precession on galaxy morphology; In section~\ref{sec:MKW}, we discuss the implication of our results, particularly on the precession of the Milky Way.

\begin{figure*}
    \includegraphics[width=\textwidth]{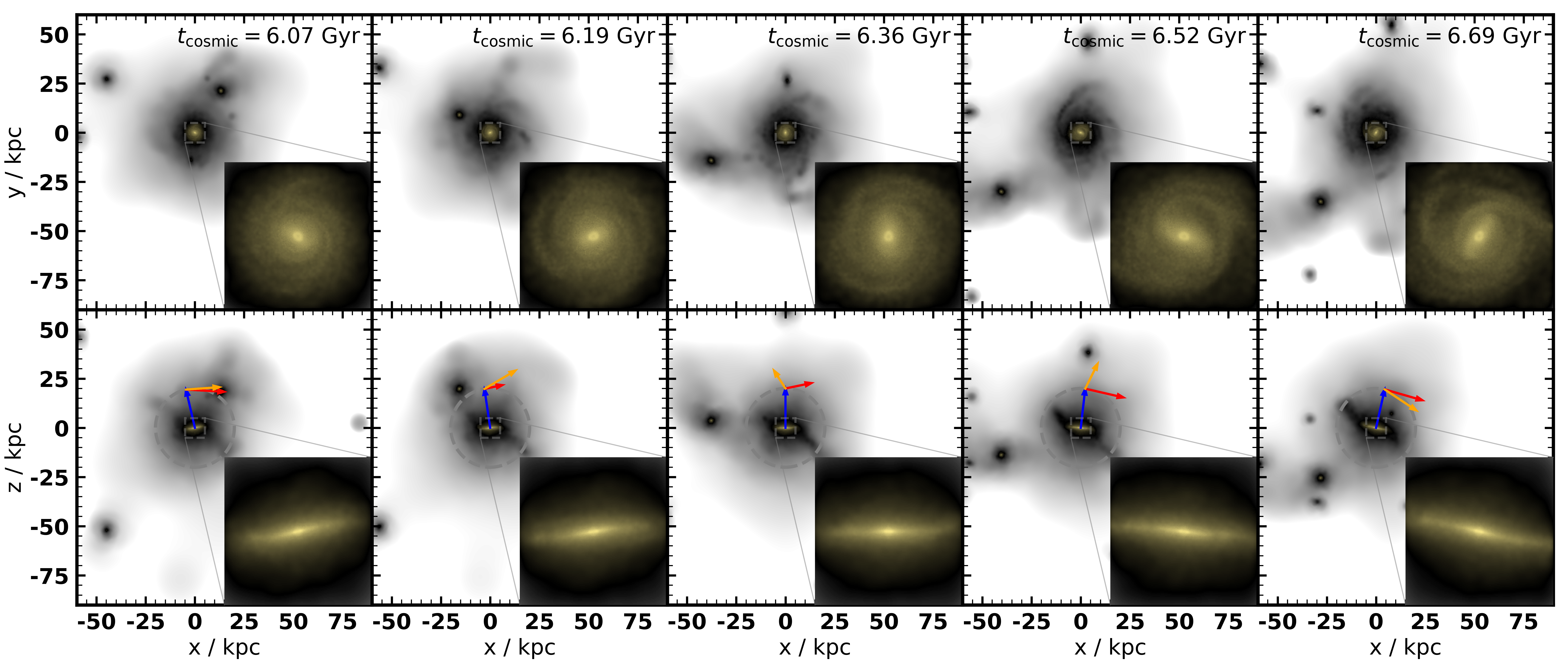}
    \caption{
Disk precession illustrated through a representative Milky Way–analog galaxy ($m_* = 1.6 \times 10^{10} \ {\rm M}_\odot$, Subhalo ID 377137) gravitationally interacting with a satellite companion ($m_* = 4.0 \times 10^{8} \ {\rm M}_\odot$; Subhalo ID 377138).
We show the projected stellar distributions in the $x$–$y$ and $x$–$z$ planes for five consecutive snapshots (cosmic times indicated in the upper panels).
All panels share the same coordinate system as that in the central column (at $t_{\rm cosmic}=6.36$ Gyr), where $z$-axis aligns with the AM vector and $x$-axis is fixed to the direction of precession. 
Blue arrows in the bottom row indicate projected stellar spin axes. 
When the arrow lies exactly in the $x$-$z$ plane, its length matches the radius of the gray dashed circle. 
Red (orange) arrows show the projected torque direction on the central galaxy from all matter within 100 kpc (the satellite companion) at that time. 
Each panel includes a 10 kpc zoom-in of the central galaxy (bottom right). Visualizations are created with Py-SPHViewer \citep{alejandro_benitez_llambay_2015_21703}.}
    \label{fig:example}
\end{figure*}

\section{Methods} \label{sec:method}

We utilize the state-of-the-art TNG50-1 hydrodynamic simulation \citep{weinberger2017Simulating, pillepich2018Simulating, springel2018First, nelson2019IllustrisTNG, Nelson2019TNG50, Pillepich2019TNG50}, which reproduces a wide range of observational results beyond its calibration set. Beyond these calibrations, it is also in good agreement with observations for other properties and statistics, such as the gas fractions and the HI mass--size relation \citep{diemer2019Atomic}, the gas-phase mass--metallicity relation \citep{torrey2019evolution}, intracluster medium properties \cite{gouin2023Soft}, and optical morphologies \citep{rodriguez-gomez2019MNRAS_optical}. 
To identify Milky Way-like analogs, we dynamically decompose galaxies into hot and cold components across their evolutionary history. Our sample is selected from central galaxies which have stellar masses between $10^{10}$ and $10^{11}\rm{M_\odot}$ at the epoch of their coldest state and maintain their dynamically cold state for at least one giga year (hereafter one-Gyr disk period). To exclude the influence of violent dynamical events, we exclude galaxies characterized by significant mergers \citep{Dodge2023} or those with a large amount of ex-situ stellar population inherited from substantial accretion events \citep{Rodriguez-Gomez2016MNRAS_stellar, sheng2024Spin}. 
In order to comprehend the influence of the procession on the later development of these disk galaxies, we additionally require that the selected galaxies do not accumulate excessive mass during their subsequent evolution. The application of these selection criteria yields a final, well-defined sample of 190 galaxies. We mainly focus on their one-Gyr disk periods and the median time of the disk period is referred to as $t_{\rm disk}$. Please see Appendix~\ref{apd:sim&sample} for the details. 

It is important to stress that major mergers and strong interactions make a substantial contribution to galactic warps, galaxy alignments, and morphological evolution studied in this work. Our decision to exclude galaxies undergoing major mergers is intended to isolate and clarify the dynamical processes governed by precession, not to understate the impact of mergers on sharping galaxy properties.

Figure~\ref{fig:example} illustrates the precession phenomenon with a representative Milky Way-like galaxy at $t_{\rm disk}\sim$ 6.36 Gyr,
gravitationally interacting with a satellite companion. 
The precession rate ($\Omega_{\rm L}$), defined as the angular velocity of the orientation of the angular momentum (AM), is $\Omega_{\rm L}=43 \deg/{\rm Gyr}$, computed as the average value 
within about 600 million years around $t_{\rm disk}$.
The precession is tightly aligned with the torques of the surrounding material within 100 kpc, suggesting that the asymmetric distribution of matter is causing the precession. 
Notably, the torque exerted by the satellite companion aligns well with the total torque, implying that massive satellites may serve as tracers of the orientation of such asymmetries \citep{arora2025Shaping}, at least for this system. 

\section{Origin of the precession of galactic disks} \label{sec:origin}

We construct a simplified model to estimate the torque $\vec{M}$ that a satellite companion exerts on a rotating annulus of the disk at radius $r$ (see Appendix~\ref{apd:theory}). While satellites alone cannot fully explain the total torque, this framework clarifies how disk precession is linked to an anisotropic mass distribution, which in our model is represented by the satellite.
Assuming the satellite has a point mass of $m_{\rm h,sat}$ for simplicity and applying a far-field approximation, the torque-induced precession rate is given by: 
\begin{equation}
\begin{split}
    \Omega_{\mathrm{M}} \approx & 15.9 \frac{m_{\mathrm{h,sat}}}{10^{11} \mathrm{M}_{\odot}} \left(\frac{R}{30 \mathrm{~kpc}} \right)^{-3} \\
    & \times \left(\frac{v_{\mathrm{c}}}{220 \mathrm{~km/s}}\right)^{-1} \frac{r}{5 \mathrm{~kpc}} \sin{2\beta} \ \mathrm{deg/Gyr},
\end{split}
\label{eq:far_field_approx0}
\end{equation}
where $R$ is the distance from the satellite to disk center, $v_c$ is the rotational velocity of the annulus and $\beta$ is the polar angle of the satellite relative to the spin axis. A similar analytical framework was previously developed to explore galactic precession driven by asymmetric matter distributions \citep{Dodge2023}. That work, however, modeled the galaxy as a whole and did not address how precession rates vary between different galactic regions. In contrast, our model reveals that the precession rate depends on radius, offering insights into the galaxy’s morphological evolution. 

\begin{figure}
    \centering
        \includegraphics[width=\columnwidth]{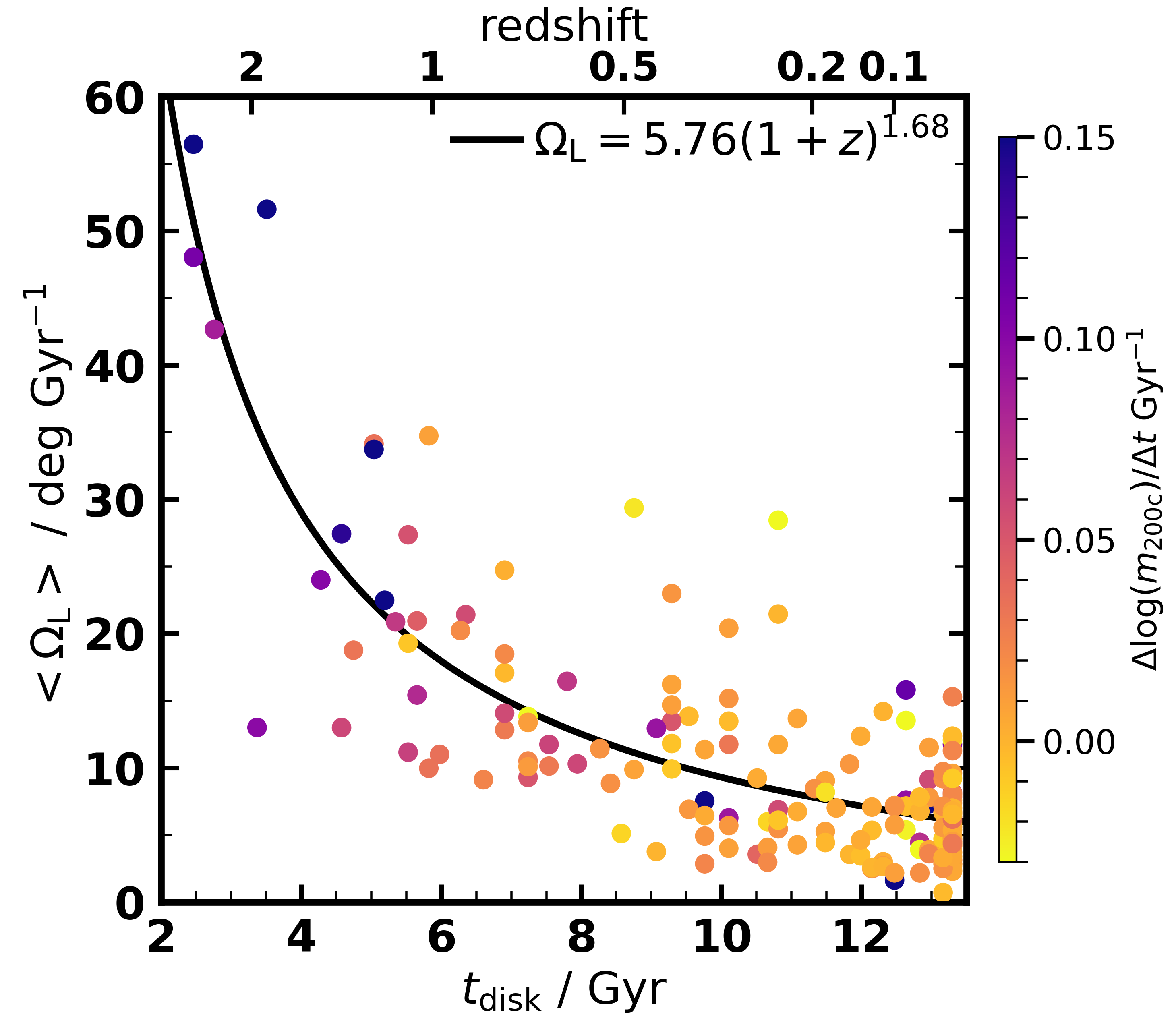}
    \caption{
    Precession rate $\Omega_{\rm L}$ as a function of $t_{\rm disk}$ for the selected galaxies with the color-coding specific halo accretion rate, defined as $\Delta\log(m_{\rm 200c})/\Delta t$. 
    The halo mass $m_{\rm 200c}$ is the mass within a spherical region, within which the mean density is 200 times the critical density.
    The precession rates, $\Omega_{\rm L}$, are calculated between consecutive snapshots during the selected disk stage and averaged over this $1 \ {\rm Gyr}$ disk period.
    The relation between $\Omega_{\rm L}$ and redshift $1+z$ is fitted using a power-law function, represented by the black line.
    }
    \label{fig:OmegaL}
\end{figure}

The precession rate decays as $R^{-3}$, indicating that the material within the central halo dominates the tidal torque \citep{semczuk2020MNRAS_Tidally}. 
Furthermore, the precession rate is proportional to the radius, establishing a radial gradient and driving differential precession that accelerates with radius assuming a flat rotation curve. This naturally generates warped structures in outer disks and accounts for the enhanced warping observed in extended HI disks compared to stellar components \citep{Bosma1981_21-cm,Briggs1990_Rules}. 
The tidal torque vanishes when the satellite lies either along the spin axis ($\beta = 0^{\circ}$ or $180^{\circ}$) or within the disk plane ($\beta = 90^{\circ}$), and reaches its maximum at $\beta \approx 45^{\circ}$. 
The configuration at $\beta=0^\circ$ (or $180^{\circ}$) represents an unstable equilibrium, where infinitesimal perturbations drive the system toward increasing $\beta$. The system ultimately converges to a stable equilibrium at $\beta=90^\circ$,  which corresponds to lying within the disk plane (see Appendix~\ref{apd:theory}).
Applying torque reciprocity between the central and the satellite, the disk precession of the satellite induced by the central galaxy and its halo is expected in a similar way. The precession of satellite disks is likely prominent due to the central halo's mass advantage ($m_{\rm h,cen}/m_{\rm h,sat} \gg 1$), potentially causing tidal locking that orients the disk edge of the satellite toward the central galaxy (see below). 

While this theoretical framework lays out the basic mechanisms driving disk precession, reproducing its behavior in a cosmological context requires simulations that include ongoing halo accretion, systems of multiple satellites, tidal deformation of dark matter halos, and the influence of the surrounding cosmic web. 
We first show the precession rate of the simulated galaxies as a function of $t_{\rm disk}$ (redshift) in Figure~\ref{fig:OmegaL}.  
The precession rate is determined by averaging the values derived from each pair of successive snapshots within the one-Gyr disk periods. 
The rate continuously increases toward higher redshifts, following a scaling of $\propto (1+z)^{1.7}$.
The rate is particularly high at $z > 1$ ($t_{\rm disk} < 6\ \mathrm{Gyr}$), reaching values up to $50 \deg/\mathrm{Gyr}$, sufficient to erase the initial orientation of the AM within a Hubble time. 
At lower redshifts, $\Omega_{\rm L}$ declines to $\sim 10 \deg/\mathrm{Gyr}$ but remains dynamically significant.
The galaxies at $z>1$ exhibit strong matter accretion, as shown in the color-coding of the halo accretion rate in Figure~\ref{fig:OmegaL}. 
This suggests that systems undergoing rapid accretion can develop a strongly anisotropic matter distribution, which in turn exerts substantial tidal torques on galaxy disks and ultimately drives strong precession, and probably causes the spin flip in dark matter halos \citep{Bett2016MNRAS_Spin_flips_II}. 
We further inspect the precession trajectories 
and reveal two distinct types: one following a linear path and the other displaying a spiral pattern, which are both smooth and continuous (see Appendix~\ref{apd:prec}).
\begin{figure*}
    \includegraphics[width=\textwidth]{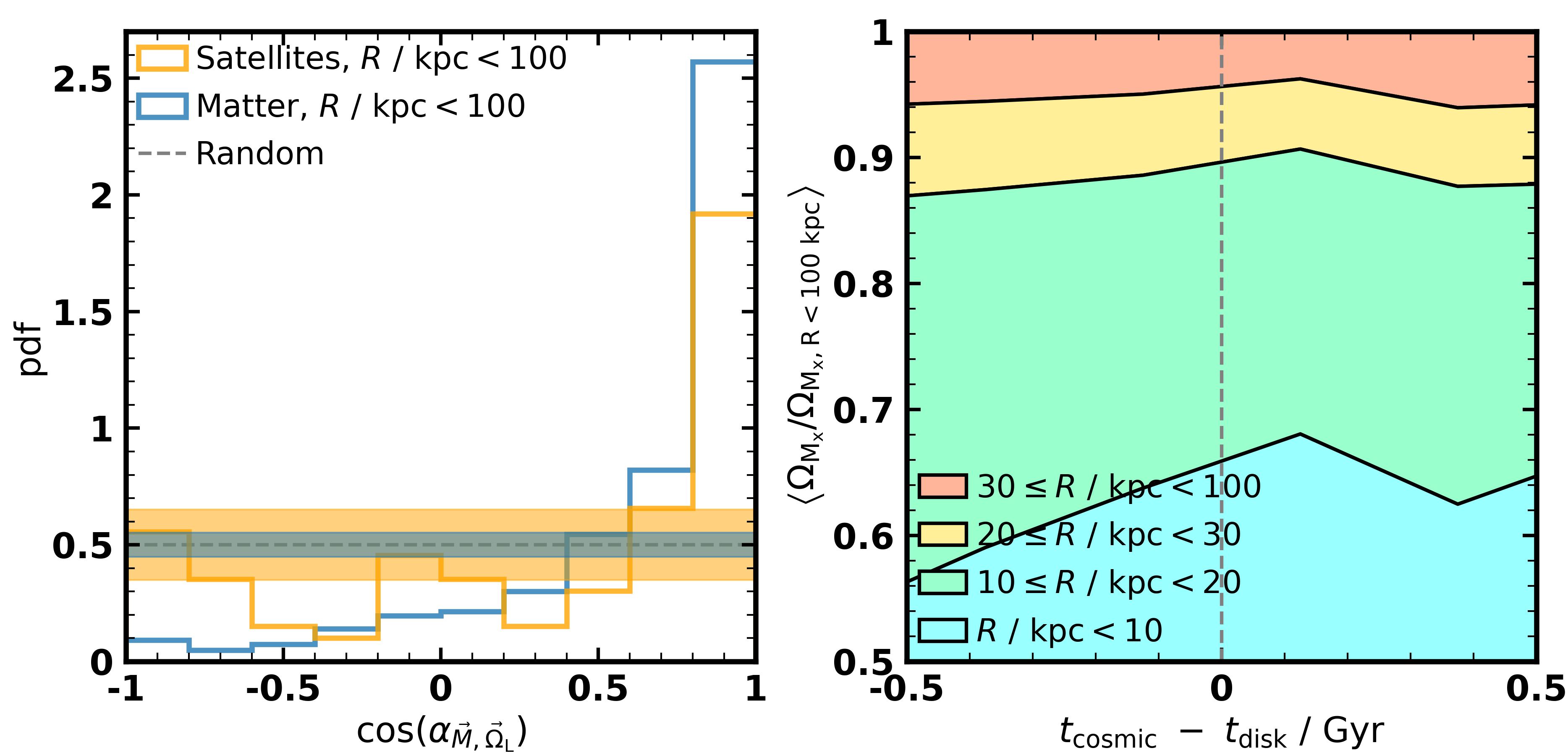}
    \caption{
    Left panel: Probability density distribution of the cosine of the angle between the tidal torque direction and the precession direction. 
    The orange line represents torques from all satellite subhalos ($m_{*, \rm sat}\geq 10^8 {\rm M}_{\odot}$) within 100 kpc, to ensure the accuracy of our measurements, we apply a selection threshold to the torque values (see Appendix~\ref{apd:prec}). The blue line shows the torques from all matter within the same radius. 
    The horizontal dashed line and shaded regions show the uniform distribution and corresponding $1-\sigma$ dispersion for reference. 
    For each galaxy,  snapshots within the one-Gyr disk period that satisfy our selection criteria are used in calculating $\alpha_{\vec{M},\vec{\Omega}_{\rm L}}$ (see Appendix~\ref{apd:prec}). Therefore, more than one $\alpha_{\vec{M},\vec{\Omega}_{\rm L}}$ for each galaxy are used. 
    The $1-\sigma$ dispersion is estimated according to the total number of $\alpha_{\vec{M},\vec{\Omega}_{\rm L}}$ used in the statistics. 
    Right panel: Relative contributions to the total torque from different radial regions.
    The horizontal axis shows cosmic time relative to $t_{\rm disk}$, while the vertical axis refers to the projected torque along the precession direction, normalized to the total torque from matter within 100 kpc.
    The colored regions show contributions from various radial ranges as indicated in the legend.}    
    \label{fig:alpha_and_OmegaMx}
\end{figure*}

The precession cannot be attributed to mergers, as significant merger systems are excluded due to strict sample selection. We note that merger systems are likely to exhibit stronger AM variation than the selected galaxies. 
Analysis of stellar spin axes reveals consistent orientation for stars formed during $t_{\rm disk} \pm 0.5$ Gyr compared to pre-existing stellar populations. 
This rules out the misalignment of newly formed stars as the origin of spin-axis reorientation and indicates an external driving mechanism.
We subsequently determine the spatial scales of the material responsible for driving the precession. 
The left panel of Figure~\ref{fig:alpha_and_OmegaMx} shows the distribution of the cosine of the angles between the precession direction and the tidal torque exerted by the surrounding material within 100 kpc (blue line). 
The strong alignment demonstrates that the precession is driven by an asymmetry in the inner halo.
The slight misalignment occurs because the tidal torque is an instantaneous measurement from a single snapshot, whereas the precession is calculated from two adjacent snapshots.
The right panel of Figure~\ref{fig:alpha_and_OmegaMx} shows the contribution to the torque of different radial regions, averaged over all samples in each time bin. 
The material within $10 \ {\rm kpc}$ dominates, the majority of which is dark matter, contributing approximately 60\% of the total torque. 
Furthermore, material within $30 \ {\rm kpc}$ accounts for almost all of the torque responsible for the precession.
This agrees with the theoretical framework in which the tidal effect scales as $\propto R^{-3}$ and thus diminishes rapidly with distance (Eq.~\ref{eq:far_field_approx0}). Moreover, the predicted precession rate ($\Omega_{\rm M}$) based on the tidal torque is on average close to $\Omega_{\rm L}$ (see Appendix~\ref{apd:prec}). All of these demonstrate that disk precession is driven by the tidal torque of local environment.

\section{Galactic warps induced by disk precession} \label{sec:warp}

\begin{figure*}
    \centering
    \includegraphics[height= 8cm]{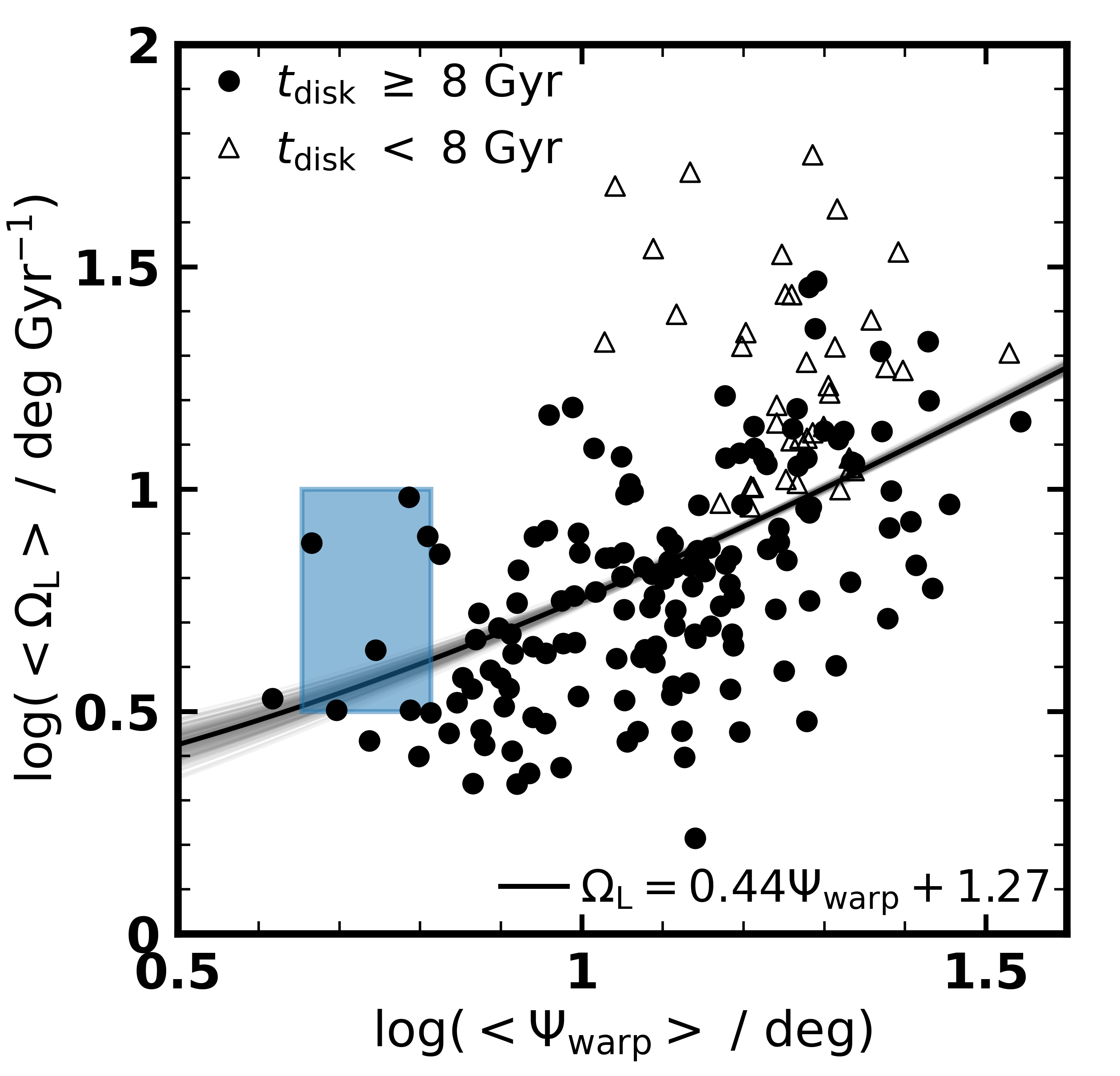}
    \includegraphics[height= 8cm]{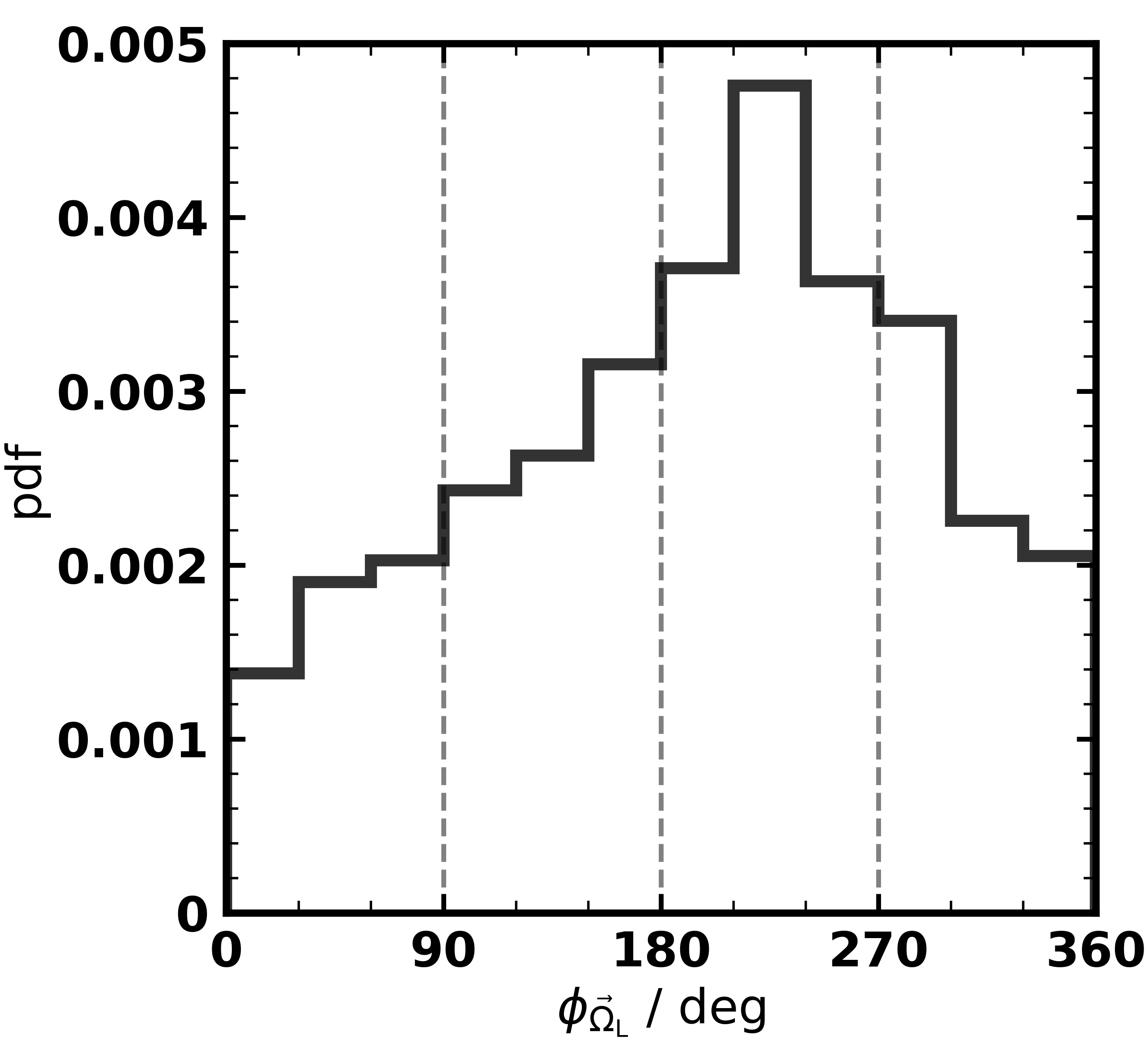}
    \includegraphics[height= 8cm]{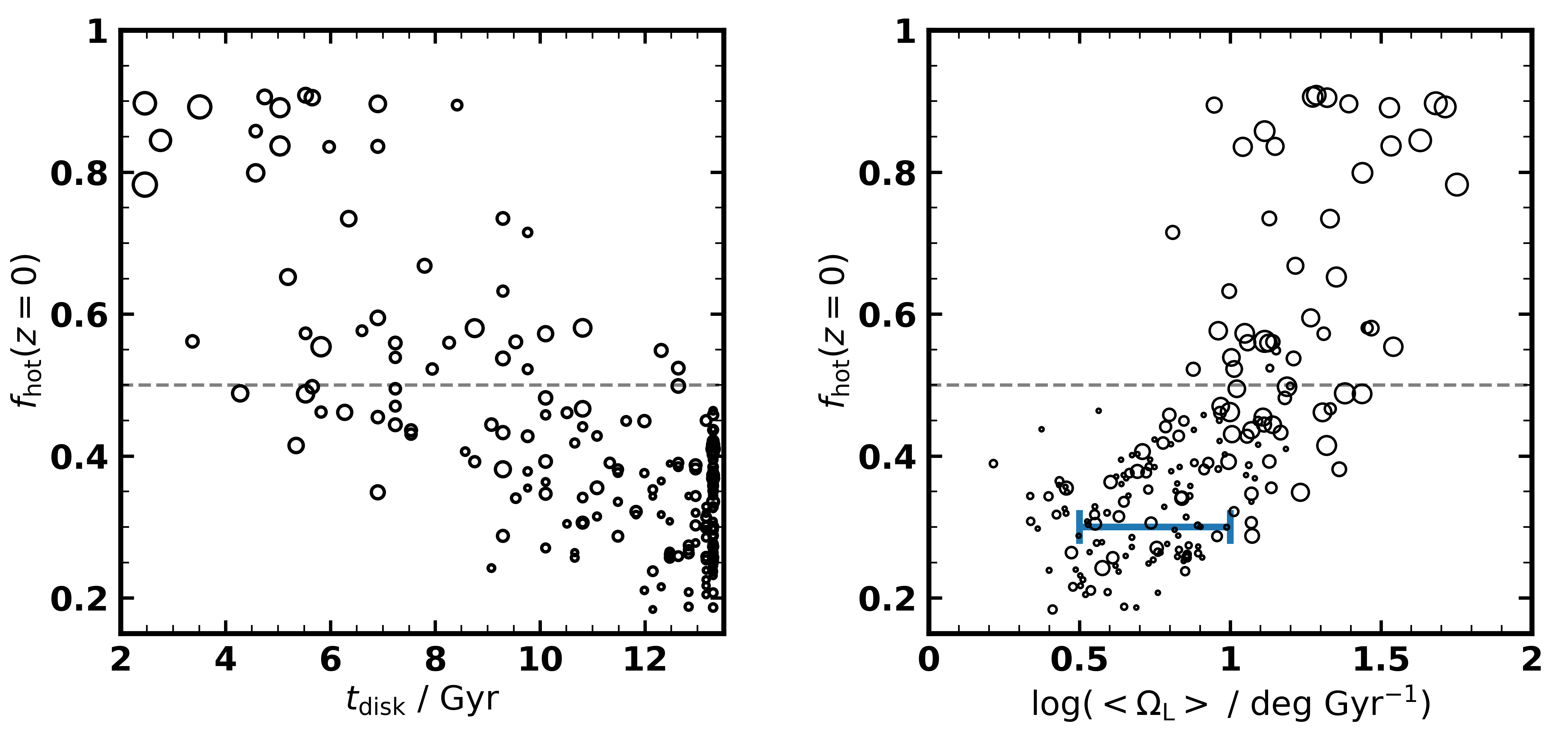}
    \caption{
    Upper left panel: The precession rate as a function of the warp angle for the selected galaxies. Both quantities are averaged over the one-Gyr disk period. The solid circles show the galaxies with $t_{\rm disk} \geq 8 \ {\rm Gyr}$ while the open triangles for galaxies with $t_{\rm disk} < 8 \ {\rm Gyr}$. 
    $\Omega_{\rm L}$ is strongly correlated with $\Psi_{\rm warp}$, particularly for disks at low redshift.
    The thick black line shows the best-fit result from a Markov chain Monte Carlo (MCMC) linear fit to the $t_{\rm disk}\ge8\rm Gyr$ sample, while the lighter lines correspond to 200 randomly selected MCMC realizations.
    The blue rectangle indicates the potential location of the Milky Way. The length of the rectangle is determined by the observational data \citep{Levine2006}, while the height indicates the range of the simulated galaxies with warp similar to that of the Milky Way.
    Upper right panel: Probability distribution of the azimuthal angle of the precession direction in a coordinate system defined by the spin vector and the warp direction (see Appendix~\ref{apd:warpangle}).
    Each snapshot within the disk period contributes one event.
    Bottom row: The relations of the hotness $f_{\rm hot}$ at $z=0$ with $t_{\rm disk}$(left) and $<\Omega_{\rm L}>$(right).
    Each data point refers to a selected disk galaxy, with its size in the left (right) panel decided by $<\Omega_{\rm L}>$ ($t_{\rm disk}$). 
    The blue line segment shows the potential location of the Milky Way, with the hotness equal to the bulge fraction of the Milky Way \citep{Quezada2025A&A} and the length indicating the range of $<\Omega_{\rm L}>$ for the corresponding simulated galaxies.
    }
    \label{fig:WarpAngle}
\end{figure*}

Now, we start investigating the impact of disk precession on the properties and orientation of galaxies.
Galactic warps, which are often observed in the Milky Way and nearby galaxies, 
have been extensively studied observationally and theoretically over the past half-century \citep{Bosma1981_21-cm, Briggs1990_Rules, reshetnikov2002AA_Statistics, Jozsa2007_Kinematic, chen2019NatAs_intuitive, Poggio2020NA_Evidence, han2023NatAs_tilted, huang2024NatAs_slightly}, yet their formation mechanisms remain debated \citep{sparke1988MNRAS_model, masset1997AA_Non-linear, debattista1999ApJ_Warped, shen2006galactic, Garcia-Conde2024AAGalactoseismology, reshetnikov2025Galactic, Sankar2025arXiv_hot}.
We quantify the warp amplitude of cold gaseous disk, by measuring the maximum bending angle ($\Psi_{\rm warp}$) \citep{Ann_warp06, Kim_warp14} within 4 half-mass radius ($4r_{\rm e}$) from galactic center (see Appendix~\ref{apd:warpangle}).  
We show the precession rate as a function of $\Psi_{\rm warp}$ in Figure~\ref{fig:WarpAngle}. 
As shown, a significant correlation between warps and disk precession is evident in galaxies at low redshifts, confirming their physical link. 
The fitting gives a result of $\Omega_{\rm L}/{\rm (deg \ Gyr^{-1})} = 0.44 \ \Psi_{\rm warp} / {\rm deg} +1.27$, and the Spearman correlation coefficient is $\rho=0.55$ with a null-hypothesis possibility of ${\rm p}=1.24\times10^{-13}$, indicating a highly significant positive correlation. 
Under the assumptions that the galactic ring behaves as a rigid body and that the torque source remains fixed, one would anticipate the warp direction to be opposite to the direction of precession.  The warp direction is defined to lie within the disk plane and be oriented from the galactic center toward the location of maximum upward warp (see Appendix~\ref{apd:warpangle}). The right panel of Figure~\ref{fig:WarpAngle} presents the distribution of the azimuthal angle of the precession direction measured relative to the warp direction. This distribution peaks at an angle slightly larger than 180$^{\circ}$, in agreement with our theoretical expectation. The small offset can be attributed to the simplifying assumptions of a rigid body ring and a static torque source. These findings indicate that disk precession is an important factor in generating the galactic warps observed in these galaxies, although additional mechanisms may also play a role.

\begin{figure*}
    \centering
        \includegraphics[width=8.55cm]{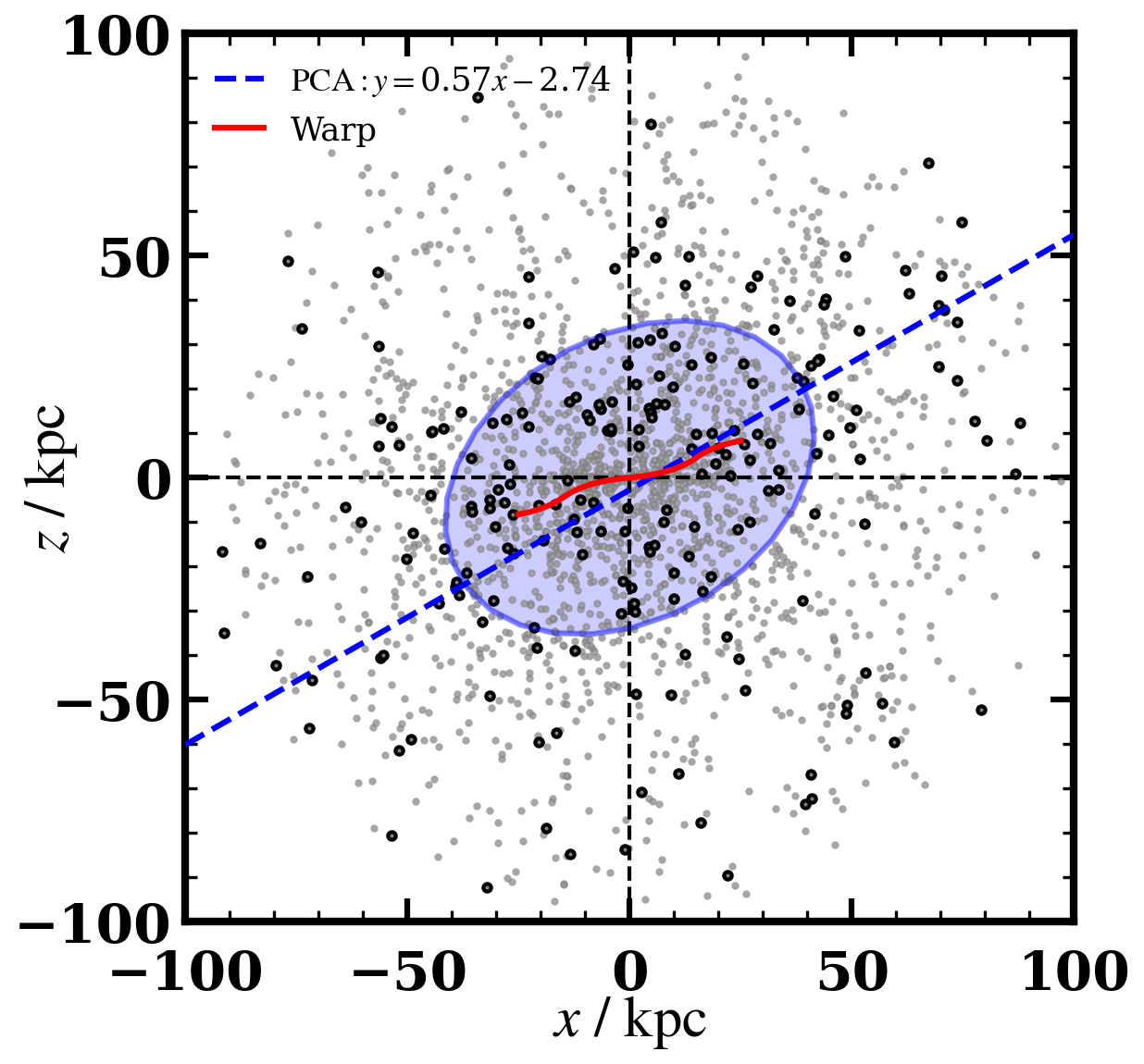}
        \includegraphics[width=8.4cm]{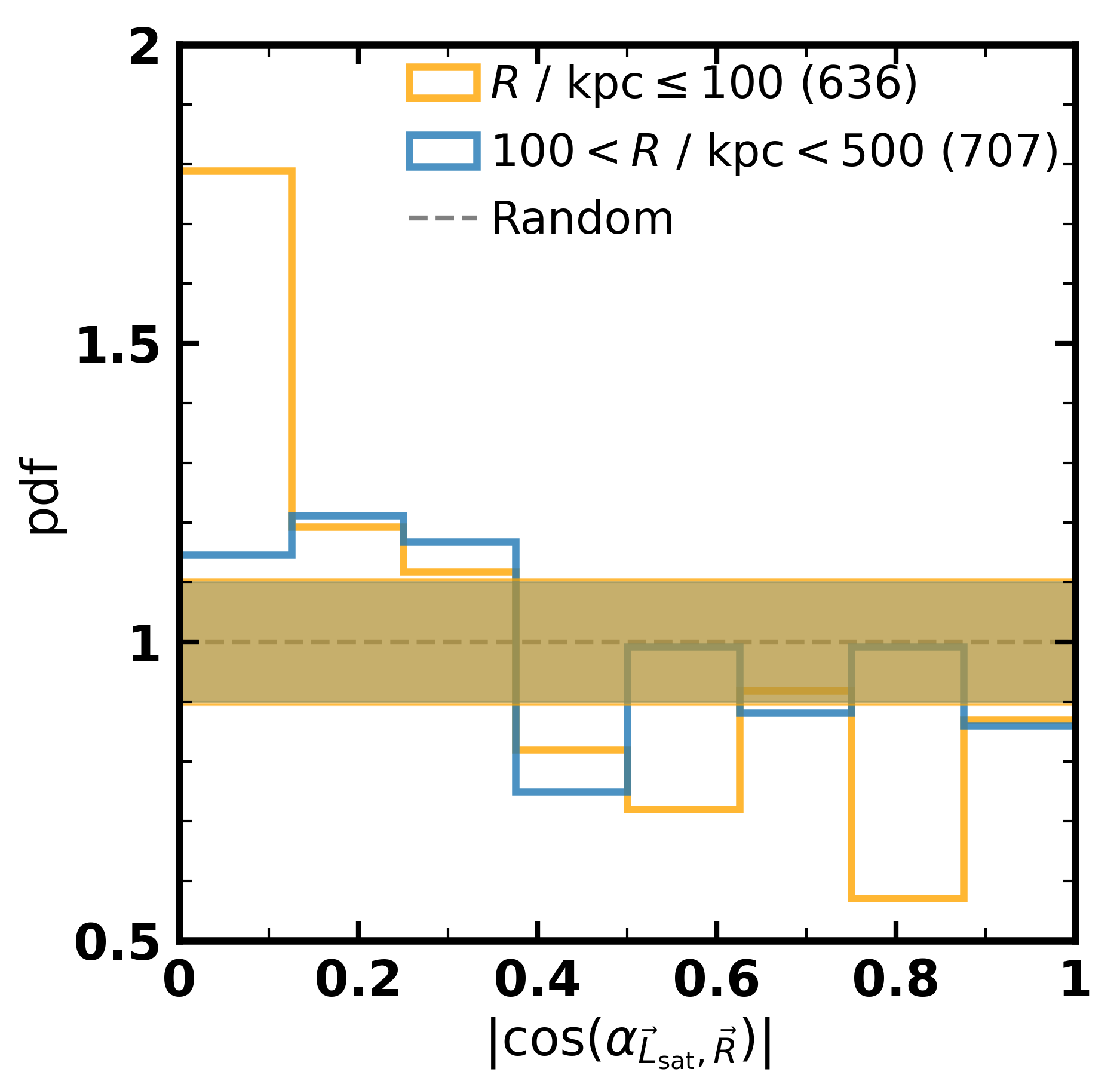}
    \caption{
Left panel: The projected spatial distribution of satellites within $100~{\rm kpc}$ to the centrals by stacking satellites around each central disk galaxy across all selected snapshots.
The size of the symbols correspond to 
the mass ratio between satellites and centrals. The $x$-axis is oriented along the warp direction, while the $z$-axis is aligned with the AM vector of the central galaxies.
The black circles represent massive satellites with $m_{*, \rm sat} > 0.05 \ m_{*, \rm cen}$. The red curve indicates the shape of warp in the stacked projection. The blue dashed line represents the principal component direction in the PCA analysis of the massive satellites. The blue ellipse corresponds to the best-fit inertia ellipsoid for the spatial distribution of massive satellite. 
Right panel: Probability distribution of the cosine of the angle between the spin vector of satellite galaxies ($\vec{L}_{\rm sat}$) and the radial vector from the central galaxy ($\vec{R}$) at $z=0$. 
The orange histogram corresponds to satellites within a radial distance of $100 \ {\rm kpc}$, while the blue one shows the result for surrounding galaxies in the $100 < R \ / \ {\rm kpc} < 500$ range.
The total counts for each region are listed in the legend.
    The horizontal dashed line and shaded regions indicate the uniform distribution and corresponding $1-\sigma$ dispersion.
Note that all satellites have $m_{\rm *, sat} > 10^{8}{\rm M}_{\odot}$.
}
   \label{fig:effects}
\end{figure*}

The torque from satellites is also strongly aligned with the precession direction (Figure~\ref{fig:alpha_and_OmegaMx}), indicating that satellites reliably trace the mass distribution responsible for disk precession. A connection is expected between warps and satellite distribution.
This expectation is borne out by observations, which find that warps are both more frequent and more evident in galaxies that have a companion or reside in a denser environment \citep{reshetnikov1998AA_Statistics, schwarzkopf2001AA_Properties}.
When our samples are stacked edge-on with the $x$-axis aligned to the warp direction, a statistical excess of satellite galaxies is found in the first and third quadrants compared to the second and fourth, as shown in the left panel of Figure~\ref{fig:effects}. The number of massive satellites in the first and third quadrants is 1.7 times the number in the second and fourth quadrants (see Table~\ref{tab:warp} in Appendix). 
This specific distribution establishes a physical connection among tidal torque, disk precession and warps. Interestingly, a recent study \cite{Zee2025arXiv_Warped} found a similar pattern for satellite distribution in observational data. Collectively, these results point to a pronounced dynamical link between galactic warps and disk precession in observational data, though we note that clearly isolating the role of precession from other, more general environmental influences remains difficult.

\section{Satellite alignment induced by disk precession} \label{sec:satellite}

The disk precession should be even more important for satellite galaxies, because the torque from central galaxies is much stronger, likely governing satellite orientations.
We tested this by analyzing the central-satellite alignment, selecting all massive central galaxies at redshift zero with $10^{10} < m_* / \mathrm{M}_{\odot} < 2 \times 10^{11}$ for robust statistics. 
The alignment is quantified by the angle $\alpha_{\vec{L}_{\rm sat},\vec{R}}$ between the satellite's spin axis and its position relative to the central galaxy, shown in the right panel of Figure~\ref{fig:effects}.
The results reveal that satellite galaxies with $R \leq 100\,\rm kpc$ tend to have AM vectors perpendicular to the direction pointing towards their centrals, so that their disks are preferentially oriented toward the central galaxy. This alignment is consistent with previous findings from both simulations \cite{lan2024Hydrodynamical} and observations \cite{pereira2005Radial, faltenbacher2007Three}. For satellites at $R > 100\,\rm kpc$, we do not observe any significant signal. This allows us to exclude large-scale cosmic tides as the main driving mechanism.
According to Equation~\ref{eq:far_field_approx0}, a state of stable equilibrium is reached when the central galaxy is situated within the plane perpendicular to the satellite’s AM vector (i.e., $\rm cos(\alpha_{\vec{L}_{\rm sat},\vec{R}}) = 0$), where the net torque vanishes. 
This strongly supports the notion that the alignment is formed due to the tidal torque after the satellites fall into the central regions of host halos, where the tidal torque becomes very strong. It is unclear whether local gravitational stretching from internal substructures could produce similar alignment effects. This alignment effect underscores its broader significance for weak lensing cosmology: the dynamical coupling between satellites and centrals represents a key source of intrinsic alignment contamination that must be accounted for to obtain unbiased cosmological constraints \citep{lee2025arXiv_effect, pandya2025arXiv_IAEmu}. 
\begin{figure*}
    \centering
        \includegraphics[width=8.6cm]{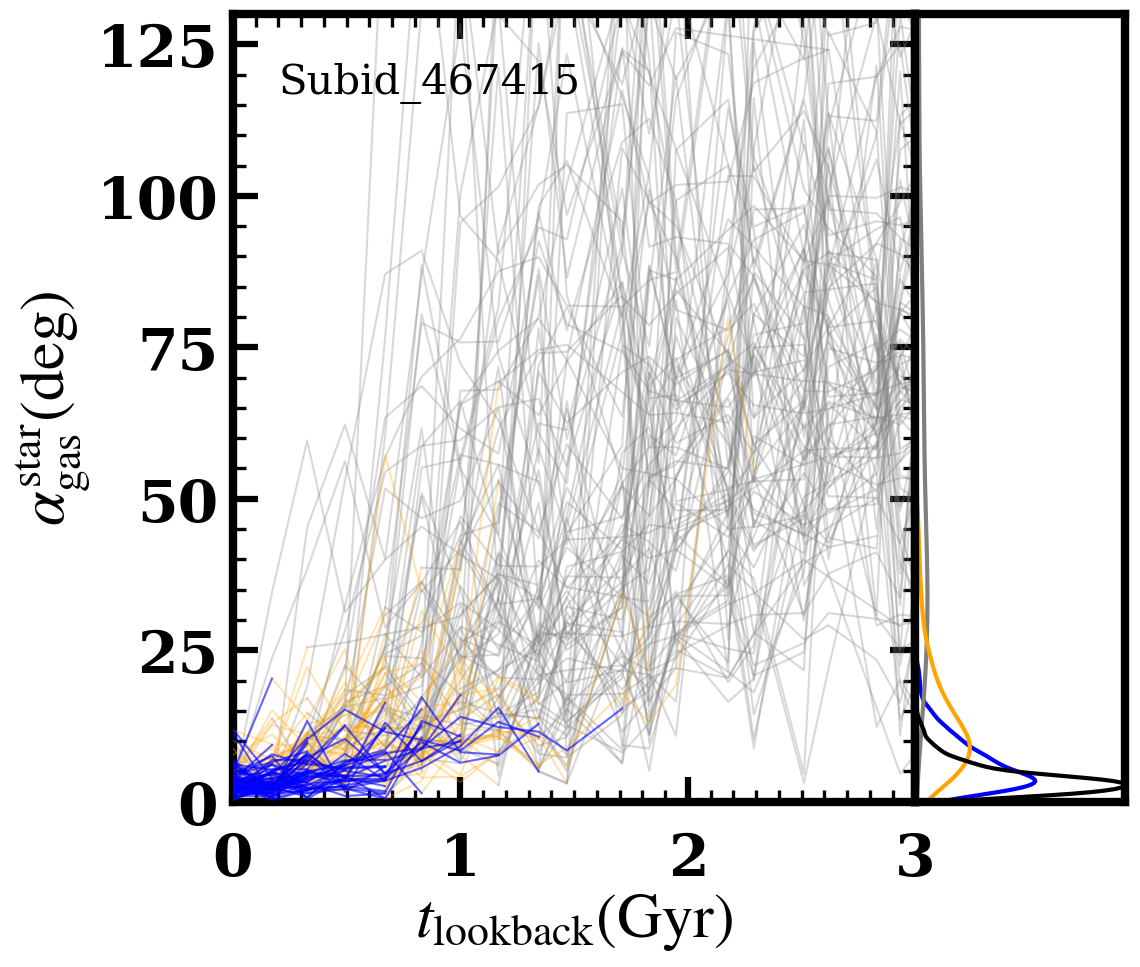}
        \includegraphics[width=8.6cm]{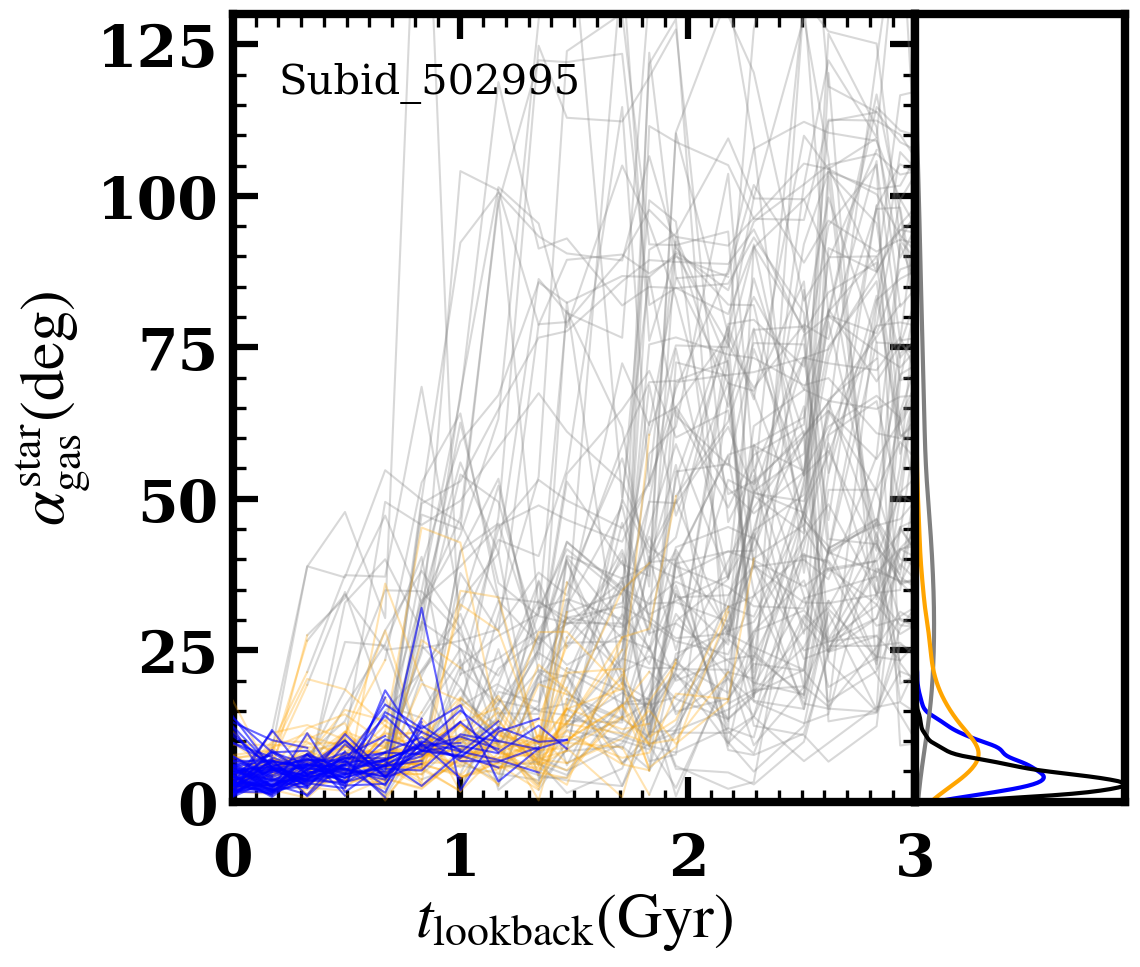}
    \caption{The evolution of the AM vectors for gas elements of two disk galaxies. We trace the evolution of the angle ($\alpha_{\rm gas}^{\rm star}$) between the AM vectors of gas elements and the spin direction of stellar component at $\rm z=0$. The left Subplot of each panel illustrates the evolutionary trajectories of gas elements that are located within the galaxies at $z=0$. The blue part of each trajectory shows the evolution of the gas from the time ($t_{\rm 2r_{\rm e}}$) when it first crossed the inner $2 r_{\rm e}$ to $z=0$, and the yellow part of each trajectory shows the evolution of the gas from the time ($t_{\rm 4r_{\rm e}}$) when it first crossed the inner $4 r_{\rm e}$ to $t_{\rm 2r_{\rm e}}$, while the grey part shows the evolution before $t_{\rm 4r_{\rm e}}$. Only 100 randomly selected trajectories are displayed for clarity. The right subplot of each panel presents the probability distributions of $\alpha_{\rm gas}^{\rm star}$ for all gas elements at four epochs: at $z=0$ (black lines), at $t_{\rm2r_{\rm e}}$ (blue lines), $t_{\rm4r_{\rm e}}$ (yellow lines) and at the time when the distances between gas elements and the galaxies first drop below $50~\text{kpc}$ (gray lines).}
   \label{fig:gas_track}
\end{figure*}

\section{Impact of precession and tidal torque on galaxy morphologies} \label{sec:morphology}

We next investigate the role of disk precession in shaping galaxy morphologies.
The morphology of simulated galaxies is often measured using the dynamically hot fraction, $f_{\rm hot}$,  the mass fraction of dynamically hot stars \citep{abadi2003Simulations, El-Badry2018MNRAS_Gas, Genel2015ApJ_Galactic}.
The bottom row of Figure~\ref{fig:WarpAngle} displays the hotness of the galaxies at $z=0$ as a function of $t_{\rm disk}$ and the mean precession rate. Disks formed earlier (smaller $t_{\rm disk}$) or that experienced stronger precession tend to be dynamically hotter now.
Previous studies also found that the velocity dispersion of stars increases monotonically with age \citep{McCluskey2024heating}. 
All of these suggest that tidal torque and precession play an important role in exchanging angular momentum between galaxies and environments and driving secular evolution by heating stellar orbits. A fraction of the selected galaxies can be classified as elliptical galaxies at $z=0$, because they have $f_{\rm hot}>0.8$, representing the dynamically hottest population in the simulation. 
These elliptical galaxies appear more prolate than elliptical galaxies that have ever experienced major mergers (see Appendix~\ref{apd:axis_ratio}). Therefore, disk precession may play an important role in the formation of elliptical galaxies with high ellipticities. Other dynamical mechanisms associated with anisotropic mass distributions are also crucial for the formation of prolate elliptical systems. In addition, minor mergers could contribute to further heating of the galaxy. More comprehensive studies are clearly required to fully assess the impact of disk precession and other processes.

The tidal torques exerted by central disk galaxies can also regulate cold gas streams in a manner analogous to their influence on satellites. As cold gas condenses and falls onto central galaxies, its AM vector is typically initially misaligned with that of the host galaxy(Figure~\ref{fig:gas_track}). During its inward spiraling motion, this AM vector is progressively reoriented by the strong torques from the central disks \citep{Danovich2015Fourphase}. 
The configuration remains stable only when the gas streams and the central disks rotate within the same plane (see Equation~\ref{eq:far_field_approx0}). Consequently, we use Monte Carlo tracer particles \citep{Genel_tracer, Nelson_tracer} to follow the evolution of the gas and track the spin of the gas streams (see Appendix~\ref{apd:angular_align}). As shown in Figure~\ref{fig:gas_track}, gas streams gradually align their AM vectors with the spin axes of the central disks, in a process reminiscent of galactic precession. 
Because the star formation timescale scales with the square root of the surface mass density \citep{ShiY2018SFlaw}, the coplanar configuration of cold gas and stellar disks promotes the onset of star formation on the disks. 
Hydrodynamic processes, such as thermal pressure gradients \citep{2015Danovich} or dynamical friction \citep{2009Dekel} between gas cooling streams, could also potentially affect the angular momentum evolution. Our analysis shows that the effect of the alignment process is very significant at 4 times $r_{\rm e}$, which are around 23-30 $\rm Kpc$ for the two galaxies. This may put an important constraint on the underlying physical mechanisms.

\section{Implication for the precession of the Milky Way} \label{sec:MKW}

We conclude that the precession of galactic disks is an ubiquitous phenomenon in the universe (Figure~\ref{fig:OmegaL}), as gravitational interactions universally influence all galaxies. 
This phenomenon is not limited to disk galaxies. Elliptical galaxies may undergo a more significant precession caused by nearby structures because their AM is much less than that of disk galaxies considered in this work.
By establishing a causal relationship between disk precession, warps, and galaxy alignment, we present testable predictions for future observational studies. Our study also sheds light on the formation of disk warps in the Milky Way \citep{Poggio2020NA_Evidence}.
For example, the HI warp angle for the Milky Way between 16 and 22 kpc 
is about $4.5\sim6.5$ degrees \citep{Levine2006}. We thus estimate a precession rate of $3\sim10$ degree/Gyr for the the Milky Way according to the $\Psi_{\rm warp}-\Omega_{\rm L}$ relation (Figure~\ref{fig:WarpAngle}). The estimation is also consistent with the $f_{\rm hot}-\Omega_{\rm L}$ relation, given that the mass of the central component for the Milky Way is about $30\%$ of its total stellar mass \citep{Quezada2025A&A}.

\begin{acknowledgments}
We thank the anonymous referee for a useful report. This work is supported by the National Natural Science Foundation of China (NSFC, Nos. 12595312, 12192224, 12473008). HYW thanks the support of CAS Project for Young Scientists in Basic Research, Grant No. YSBR-062 and the New Cornerstone Science Foundation through the XPLORER PRIZE. 
ECW thanks the support of the Start-up Fund of the University of Science and Technology of China (No. KY2030000200).
FM gratefully acknowledges funding from the European Union – NextGenerationEU, in the framework of the HPC project – “National Centre for HPC, Big Data and Quantum Computing” (PNRR – M4C2 – I1.4 – CN00000013 -– CUP J33C22001170001).
The authors gratefully acknowledge the support of Cyrus Chun Ying Tang Foundations. 
The work is supported by the Supercomputer Center of University of Science and Technology of China.
\end{acknowledgments}

\appendix
\vspace{-4ex}
\section{Simulation and Sample selection} \label{apd:sim&sample}
In this work, we take advantage of the publicly available TNG50-1 simulation, the highest resolution realization of the IllustrisTNG suite \citep{weinberger2017Simulating, pillepich2018Simulating, springel2018First, nelson2019IllustrisTNG}, for the reason that its galaxy morphologies conform well to observations \citep{gong2025Mock}. 
It is a cosmological magnetohydrodynamic (MHD) simulation in the $\Lambda \rm CDM$ framework along with baryonic processes that are important for galaxy formation and evolution, such as radiative cooling, star formation, chemical enrichment, stellar feedback, and active galactic nucleus (AGN) feedback. 
The galaxy formation model is calibrated to broadly reproduce several observation results: the history of star formation rate density, the stellar mass function for galaxies, the stellar mass--halo mass and stellar mass--stellar size relations, the halo gas fraction, and the black hole--galaxy mass relation. 
There are still limitations in the adopted subgrid prescriptions, and discrepancies persist when these are compared with alternative subgrid implementations. However, these uncertainties can at most reduce the precision of our results rather than invalidate our main conclusions, since our analysis is primarily concerned with gravitational interactions and dynamical processes in Milky-Way-mass galaxies, which are largely decoupled from the details of subgrid baryonic physics.

The adopted cosmological parameters are based on the results from the Planck Collaboration in 2015 \citep{planck2016Cosmological},: $\Omega_{\rm m}=0.3089$, $\Omega_{\rm b} = 0.0486$, $h=0.6774$, and $\sigma_8=0.8159$. 
TNG50-1 has a comoving box size of $35 \ {\rm Mpc/}h$, and a mass resolution of $m_{\rm dm}=3.1\times10^{5}\ {\rm M}_{\odot}/h$ and $m_{\rm baryon}=5.7\times10^{4}\ {\rm M}_{\odot}/h$. 
The publicly available \texttt{FOF} group (halo) catalogs, \texttt{S\footnotesize{UBFIND}} subhalo (galaxy) catalogs and \texttt{S{\footnotesize UB}L{\footnotesize INK}} merger trees are also used to select sample galaxies and trace their histories. 
All values in this work are reported in physical units.
To identify disk galaxies, we follow previous works \cite{abadi2003Simulations, Genel2015ApJ_Galactic, El-Badry2018MNRAS_Gas}, and compute the circularity parameter of each stellar particle. The parameter is defined as the ratio of the particle’s AM component along the total AM direction to that of a circular orbit with the same binding energy:
\begin{equation}
    \epsilon_{\rm c}=\frac{j_{\rm z}}{j_{\rm c}(E)}.
\end{equation}
The center of a galaxy is the location with the minium gravitational potential in the subhalo, and the galaxy velocity is
calculated using all star particles within twice the half stellar mass radius ($r_{\rm e}$). 
The specific energy $E$ and the specific AM in a circular orbit $j_{\rm c}(E)$ are obtained following the reference \cite{El-Badry2018MNRAS_Gas}.
We define galaxy hotness as the fraction of the stellar mass with circularity $\epsilon < 0.7$ within $2r_{\rm e}$, $f_{\rm hot} \equiv m_{*,\epsilon_{\rm c} < 0.7} / m_{*} $, where $m_*$ is the total stellar mass enclosed within the same radius. This parameter correlates strongly with morphology: cold systems ($f_{\rm hot} < 0.5$) exhibit disk-dominated structures, while hot systems ($f_{\rm hot} > 0.8$) correspond to spheroidal and elliptical galaxies.
We further define the time of maximum disk fraction, $t_{\rm max}$, as the epoch when the disk fraction $f_{\rm disk} \equiv 1 - f_{\rm hot}$ reaches its peak. 

We aim to investigate the Milky-Way-like galaxies formed at different cosmic epochs. 
Accordingly, our sample galaxies are chosen to be central galaxies within the mass range of the Milky Way at $t_{\rm max}$, and to be able to maintain its disk morphology for at least $1 \ {\rm Gyr}$ around this time. 
Its stellar mass at $t_{\rm max}$ is also required to be greater than half of the maximum stellar mass in history, to investigate their secular evolution until redshift zero. 
Here are our primary selection criteria:
\begin{equation}
    \begin{gathered}
        10^{10}\leq m_{*}(t_{\rm max}) \ / \ {\rm M}_\odot < 10^{11}, \\
        m_{*}(t_{\rm max})>\max{\{m_{*}(t)\}}\ / \ 2, \\
        \left. {f_{\rm hot}({t})<0.5} \right\vert_{ \ t_{\rm disk}-{\rm 0.5Gyr}<t<t_{\rm disk}+{\rm 0.5Gyr}},\\
    \end{gathered}
\end{equation}
where the disk time, $t_{\rm disk}$, is defined as the midpoint of the 1 Gyr interval during which the galaxy continuously satisfies $f_{\rm hot} < 0.5$ 
around $t_{\rm max}$. This is referred to as one-Gyr disk period, corresponds to $t_{\rm disk} \pm 0.5~{\rm Gyr}$.
Note that $t_{\rm disk}$ is not necessarily equal to $t_{\rm max}$. A smaller $t_{\rm disk}$ generally indicates earlier disk formation in cosmic history.
After applying these criteria, 209 galaxies are selected. 

We then minimize the effects of mergers, which can drastically change the spin and morphology of galaxies. Our selection criteria are: (1) no major merger (merger ratio $\eta>0.1$) has occurred since disk formation, and (2) the ex-situ stellar mass ratio at $z=0$ is less than 0.1. Here, $\eta$ is defined as the maximum stellar mass of the merging galaxy divided by the central galaxy's stellar mass just before the merger. The ex-situ stellar mass $m_{\rm *, ex}$ is the total mass of stars formed outside the main progenitor branch \citep{Rodriguez-Gomez2016MNRAS_stellar}. The above two criteria can be written as:  
\begin{equation}
    \begin{gathered}
        \frac{m_{*,ex}(z = 0) - m_{*,ex}(t_{\rm disk}-0.5\rm Gyr)}{m_{*}(z =0)}<0.1, \\
        \max\limits_{t>t_{\rm disk}-0.5}\{\eta(t)\}<0.1.
    \end{gathered}
\end{equation}
After applying these criteria, we obtain a sample of 190 Milky-Way-like galaxies for the analysis of disk precession.

\section{Precession rate and torque} \label{apd:prec}
The precession rate is defined as :
\begin{equation}
\vec{\Omega}_{\rm L} \equiv d\vec{L}_ {\rm tan}/Ldt,
\end{equation}
and describes the angular velocity of the stellar spin axis.
In this expression, $\vec{L}$ is the total stellar AM, calculated using all star particles contained within $2r_{\rm e}$, and $L = |\vec{L}|$. The quantity $d\vec{L}_ {\rm tan}$ denotes the tangential component of the change in $\vec{L}$, perpendicular to $\vec{L}$, and its direction sets the direction of precession.

\begin{figure*}
   \centering
   \includegraphics[width=\columnwidth]{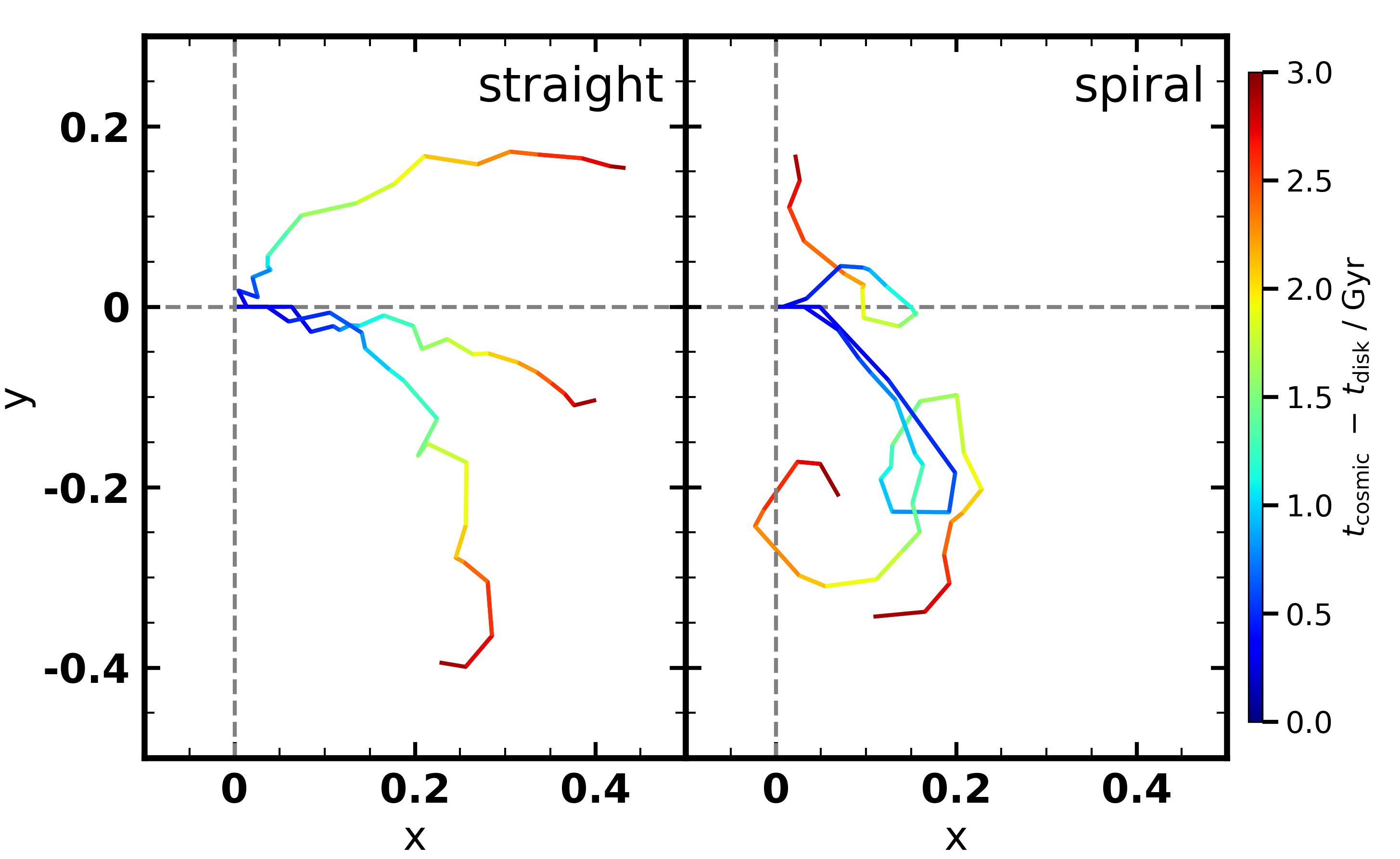}
\caption{ Projected trajectories of spin vectors for six selected typical galaxies. 
The $z$-axis($x$-axis) of the coordinate system is aligned with the spin vector (precession direction) of a galaxy at $t_{\rm disk}$.
Each trajectory shows the evolution of $(x,y)$ components  of the unit spin vector for a galaxy, color-coded by the relative cosmic time. 
Only the evolution between $t_{\rm disk}$ and $t_{\rm disk}+3 \rm Gyrs$ is displayed.
Therefore, at the starting point ($t_{\rm disk}$), both $x$ and $y$ components are zero.
The two panels show the two representative categories of trail as denoted by ``straight" and ``spiral", respectively. 
}
    \label{fig:precession_trajectory}
\end{figure*}

Due to the discrete nature of the simulation snapshots, it is not possible to obtain instantaneous precession rates. Instead, we estimate them using two adjacent snapshots.
For a given snapshot $i$, we construct a coordinate system where the $z$-axis is aligned with the stellar spin direction, $\hat{z} \parallel \vec{L}_i$. The $y$-axis is defined to be orthogonal to the spin vectors of snapshots $i$ and $i+1$, such that $\hat{y} \parallel \vec{L}_i \times \vec{L}_{i+1}$. Consequently, the $x$-axis lies in the direction of precession, $\hat{x} = \hat{y} \times \hat{z}$. The precession vector is then computed as follows:
\begin{equation}
    \begin{gathered}
        \vec{\Omega}_{\rm L_{i}} = \frac{\alpha_{\vec{L}_{i},\vec{L}_{i+1}}}{t_{i+1}-t_{i}} \hat{x},
    \label{eq:omegaL}
    \end{gathered}
\end{equation}
where $\alpha_{\vec{L}_{i},\vec{L}_{i+1}}$ denotes the angle between the spin vectors at snapshots $i$ and $i+1$, and $t_{i+1} - t_{i}$ is the corresponding time interval.
The time-averaged precession rate over the one-Gyr disk period shown in Figures~\ref{fig:OmegaL} and \ref{fig:WarpAngle} is calculated by summing the precession angles ($|\alpha_{\vec{L}_{i},\vec{L}_{i+1}}|$) of all intermediate snapshot pairs 
and dividing by the total elapsed time ($\sim 1\rm~Gyr$). 

\begin{figure*}
    \centering
    \includegraphics[width= 10cm]{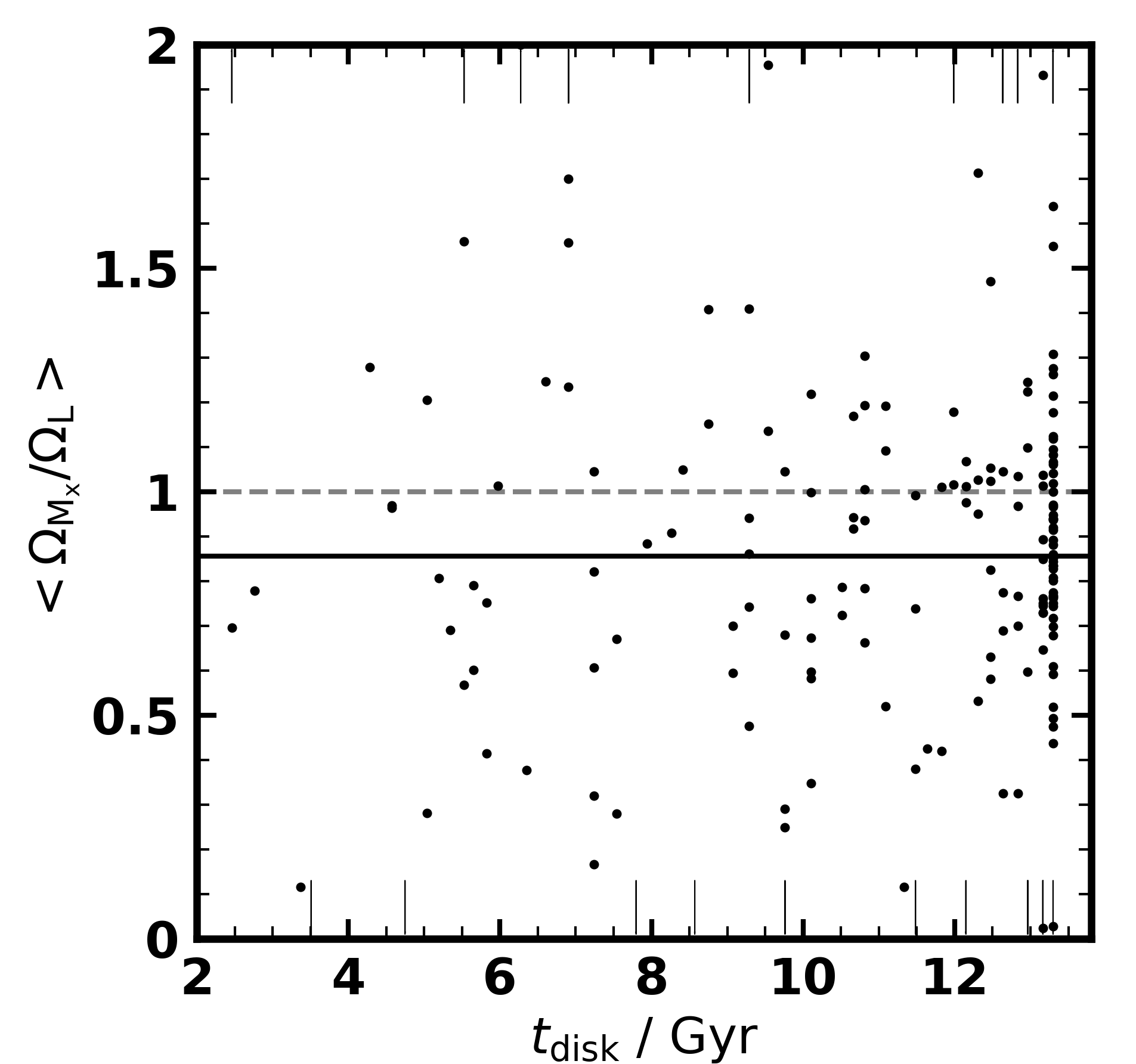}
    \caption{The relation between $<\Omega_{\rm M_x}/\Omega_{\rm L}>$ and $t_{\rm disk}$. $<\Omega_{\rm M_x}/\Omega_{\rm L}>$ is the averaged ratio of the projected torque-induced precession rate ($\Omega_{\rm M_x}$) to the precession rate ($\Omega_{\rm L}$).
    The results are averaged over the disk period. The short line segments at the edges denote outliers. 
    The median value of all these galaxies, close to unity, is shown by the horizontal solid line.
    }
    \label{fig:OmegaM}
\end{figure*}

To trace the trajectories of disk precession, in Figure~\ref{fig:precession_trajectory} we show the trail of the unit spin vector projected on the plane perpendicular to the spin vector at $t_{\rm disk}$.
Each solid line refers to the evolution of a galaxy over a $3 \ {\rm Gyr}$ period starting from $t_{\rm disk}$, colored by cosmic time relative to $t_{\rm disk}$. The trajectories are 
smooth and continuous.
We found two representative types of trail denoted as ``straight'' and ``spiral" in the two panels. For the straight pattern, the precession direction is almost unchanged. This indicates a tidal torque with a relatively stable direction. In the spiral pattern, the direction of the trajectories changes rapidly, indicating a complex and dynamically evolving environment. Note that the spiral type constitutes the majority of our sample galaxies.

\begin{figure*}
    \centering
    \includegraphics[width=\columnwidth]{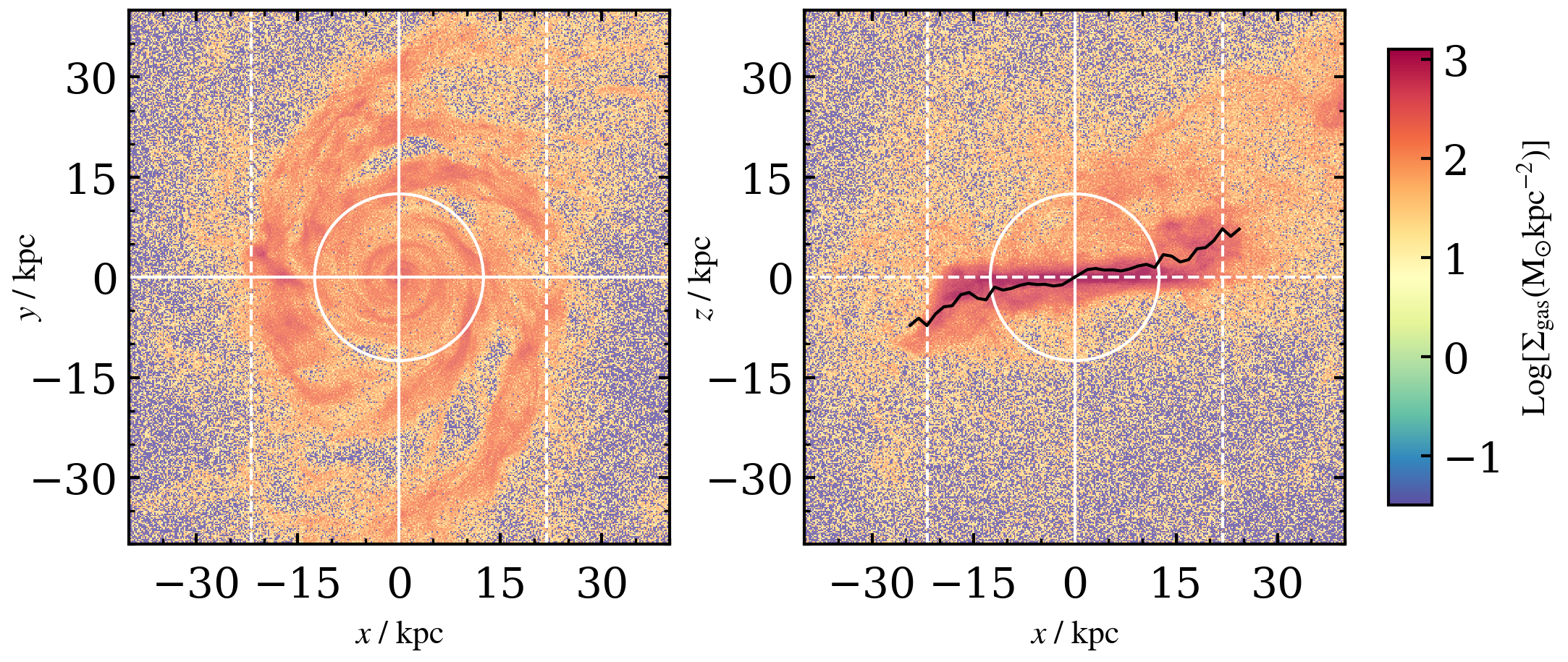}
    \caption{Cold gas warp for a galaxy (Subid 482892 in Snap 82). Left panel shows the density map in face-on projection and right panel shows the map in edge-on projection. The $x$ axis aligns with the warp direction (see Appendix~\ref{apd:warpangle}). The white cycle indicate $2r_{\rm e}$, black curve in the right panel indicates the S-type warp. The vertical white dashed lines indicate the radii where the maximum warp angle was evaluated.} 
    \label{fig:warp_proj}
\end{figure*}

We also calculate the torques $\vec{M}$ induced by the surrounding matter or satellite subhalos, in order to investigate the origin of precession and warp. 
The total torque exerted on the disk is calculated by $\vec{M}=\sum_{i}\sum_{j} \vec{r}_j \times \vec{F}_{ij}$, where the index $i$ is enumerated for all particles from the inspected torque source and $j$ for all galactic star particles within $2r_{\rm e}$. 
The $\vec{r}_j$ is the position of disk particle j relative to the galactic center, and $\vec{F}_{ij}$ is the gravitational force on particle $j$ from source particle $i$.
Due to the enormous number of particles, we employed the Tree algorithm to optimize the computation.
The torque-induced precession rate can be estimated by $\Omega_{\rm M_{\rm x}} = {M}_{\rm x} / L$, where ${M}_{\rm x}$ is the component of the torque projected in the direction of the procession and $L$ is the stellar angular momentum.

Figure~\ref{fig:OmegaM} shows the ratio $\Omega_{\rm M_x}/\Omega_{\rm L}$, calculated using all matter within a radius of $100 \ {\rm kpc}$ and averaged over snapshots within disk period.
The horizontal solid line represents the median value across all galaxies, which is close to unity ($\sim 0.9$), indicating that the tidal torques exerted by matter within $100 \ {\rm kpc}$ can almost fully account for the observed precession.
However, the large dispersion suggests that
the limited temporal resolution of the simulation snapshots makes it difficult to capture this correspondence accurately when simultaneously using the instantaneous torques and the precession rates estimated by two adjacent snapshots.

In addition to the torque amplitudes, we also examine their orientations, quantified by the angle between the torque vector and the precession direction, $\alpha_{\vec{M},\vec{\Omega}_{\rm L}}$.
The left panel of Figure~\ref{fig:alpha_and_OmegaMx} presents the probability distribution of $\cos(\alpha_{\vec{M},\vec{\Omega}_{\rm L}})$ for two cases: torques exerted by all matter within $100 \ {\rm kpc}$ (blue) and by all satellite subhalos with $m_{*, \rm sat} \geq 10^8 \ {\rm M}_{\odot}$ (orange).
These quantities are calculated for each snapshot throughout the one-Gyr disk period.
For the alignment analysis, we further apply the following selection criteria to the precession rate $\Omega_{\rm L}$ and the torque $\vec{M}$ at each snapshot:
\begin{equation}
    \begin{gathered}
        \Omega_{\rm L} \geq 5 \ \deg {\rm Gyr}^{-1}, \\
        \frac{|\vec{M}|}{m_{*} r_{\rm e}^2} \geq 50 \ \deg {\rm Gyr}^{-2},
    \end{gathered}
\end{equation}
where $m_{*}$ is the stellar mass of the central galaxy in consideration. This helps reduce the impact of the timescale inconsistency between the measured torque and the precession rate.
It reveals a clear alignment between the directions of torques and precession for both satellite subhalos and the surrounding matter. The alignment with the total matter confirms that the precession is driven by torques from the asymmetric mass distribution, while the alignment with satellite subhalos suggests that the distribution of the satellites trace well the asymmetry of mass distribution.

\section{Disk warp angle} \label{apd:warpangle}
We investigate the role of precession and tidal forces by analyzing the warp features of their cold gas disks. 
Observationally, galactic warps are generally classified into two types: S-type and U-type.
S-type warps bend in opposite directions on either side of the disk, while U-type warps bend in the same direction. Since U-type warps are weak or absent in our sample, we focus exclusively on the S-type.
The method for measuring warps follows the approach adopted in previous studies \cite{Ann_warp06,Kim_warp14} and is briefly described below.

We first define a cylindrical coordinate system centered on the galaxy, with the $z$-axis aligned with the direction of its AM.
Next, the cylindrical radial distance $r$ is divided into 20 linear bins extending to $4r_{\rm e}$, each with a vertical thickness of $\pm2r_{\rm e}$.
For each radial bin, the azimuthal angle $\theta$ is further divided into 360 uniformly distributed directions with an opening angle of $30^{\circ}$, and the vertical displacement $z(r,\theta)$ is defined as the mass-weighted mean height of the cold gas particles (with temperature less than $10^5 \ {\rm K}$) along the $z$-direction.
Only bins with radii greater than $1.5r_{\rm e}$ and containing a sufficient number of cold gas particles are included in the analysis.
The S-type warp strength in each bin is quantified as half of the vertical displacement relative to its diametrically opposite position, which can be written as: 
\begin{equation}
    \begin{gathered}
        z(r,\theta) = \sum_{\rm i}m_{i}z_{\rm i}/\sum_{\rm i}m_{\rm i},\\
        S_{\rm warp}(r,\theta) = |z(r,\theta)-z(r,\theta+\pi)|/2.
    \end{gathered}
    \label{eq:warp_cal}
\end{equation}
The warp angle is then defined as the maximum value of the warps divided by the corresponding radius:
\begin{equation}
    \begin{gathered}
        \Psi_{\rm warp} =\max\limits_{r,\theta}\{\arctan[S_{\rm warp}(r,\theta)/r]\}.
    \end{gathered}
    \label{eq:warp_sta}
\end{equation}


We also define a warp direction, which lies within the disk plane and is oriented from the galactic center toward the location of maximum warp with $z(r, \theta)>0$, i.e the upward warp. 
Figure~\ref{fig:warp_proj} shows the spatial distribution of cold gas for a galaxy, in which the vertical axis aligns with the AM vector and the horizontal axis aligns with the warp direction.  A clear warp feature is present.
The vertical dashed lines indicate the positions of the maximum warp, demonstrating that this method can accurately identify the regions where the cold gas disk is warped most strongly. 

The relation between the precession rate $\Omega_{\rm L}$ and the warp angle $\Psi_{\rm warp}$ (both values are averaged over the disk period) is shown in Figure~\ref{fig:WarpAngle}.
They are generally linearly correlated, suggesting that the warp arises from the sustained precession of the disk. Figure~\ref{fig:WarpAngle} also shows the distribution of the azimuthal angle of the precession direction in the coordinate system with the warp direction defined as the $x$ axis and the spin vector defined as the $z$ axis. The peak of the distribution is slightly larger than 180$^{\circ}$, and can be explained by the scenario that  warps are driven by disk precession.
Considering that the precession is driven by torques from surrounding matter and satellites, we expect an asymmetric satellite distribution with respect to the warp direction.
In the left panel of Figure~\ref{fig:effects}, satellite galaxies around the sample galaxies at different epochs  are stacked, projected edge-on with the $x$-axis aligned with the warp direction.
A clear trend emerges: satellites exhibit a noticeable excess in the first and third quadrants compared to the other two, consistent with the observational results \cite{Zee2025arXiv_Warped}.

To quantify this phenomenon, we consider all satellites within a 100 kpc radius around each sample galaxy. 
To increase the statistics, we extend the investigation to all snapshots when $f_{\rm hot}<0.5$.
After projecting them along the warp direction as described previously, we compute the number ratio, $\rm EX = N_{1,3}/N_{2,4}$, where $N_{1,3}$ and $N_{2,4}$ represent the number of satellites in the first and third quadrants, and the second and fourth quadrants, respectively.
The results are summarized in Table~\ref{tab:warp}, with different columns corresponding to different lower limits of the satellite-to-central stellar mass ratio ($m_{\rm *,sat}/m_{\rm *,cen}$). The ratio $\rm EX$ reaches a significant value of approximately 1.67 for $\log m_{\rm *,sat}/m_{\rm *,cen}>-1.3$. It decreases as the mass limit decreases, which can be attributed to the inclusion of diffusely distributed objects that contribute minimally to the torques.
To assess statistical significance, we calculate the probability that a ratio greater than $\rm EX$ would occur under the assumption that the satellites would be uniformly distributed in the four quadrants.
As shown in the last row of Table~\ref{tab:warp}, the low probability values reject the hypothesis of uniform-distribution and indicate that satellites are strongly correlated with the galactic warp.
This specific distribution of satellite galaxies is direct and compelling evidence of precession causing galactic warping in simulations. 

\begin{deluxetable*}{cccc} 
\tablecaption{The numbers of satellites in different quadrants for different lower limit of $m_{\ast, \mathrm{sat}}/m_{\ast, \mathrm{cen}}$. \label{tab:warp}}
\tablewidth{0pt}
\tablehead{
\colhead{$m_{\ast, \mathrm{sat}}/m_{\ast, \mathrm{cen}}$} & 
\colhead{$>10^{-1.3}\ (5\%)$} & 
\colhead{$>10^{-2}\ (1\%)$} & 
\colhead{$>10^{-2.5}\ (0.3\%)$}
}
\startdata
$N_{1,3}$ & 132 & 813 & 1171 \\
$N_{2,4}$ & 79 & 583 & 866 \\
$\mathrm{EX} = N_{1,3}/N_{2,4}$ & 1.67 & 1.39 & 1.35 \\
$p(\geq \mathrm{EX})$ & $1.6 \times 10^{-4}$ & $4.1 \times 10^{-10}$ & $7.5 \times 10^{-12}$ \\
\enddata
\tablecomments{$N_{1,3}$ represents the total number of satellites in the first and third quadrants within $100~\mathrm{kpc}$ and $N_{2,4}$ represents the total number of satellites in the second and fourth quadrants. $\mathrm{EX}$ is the ratio of them. $p(\geq \mathrm{EX})$ represents the probability of the ratio to be $\geq \mathrm{EX}$ for a randomly distributed mock distribution, where satellites are placed isotropically and uniformly.}
\end{deluxetable*}

\clearpage 
\onecolumngrid 

\section{Axis ratios for dynamically hot galaxies} \label{apd:axis_ratio}
\begin{figure*}
   \centering
   \includegraphics[width=10cm]{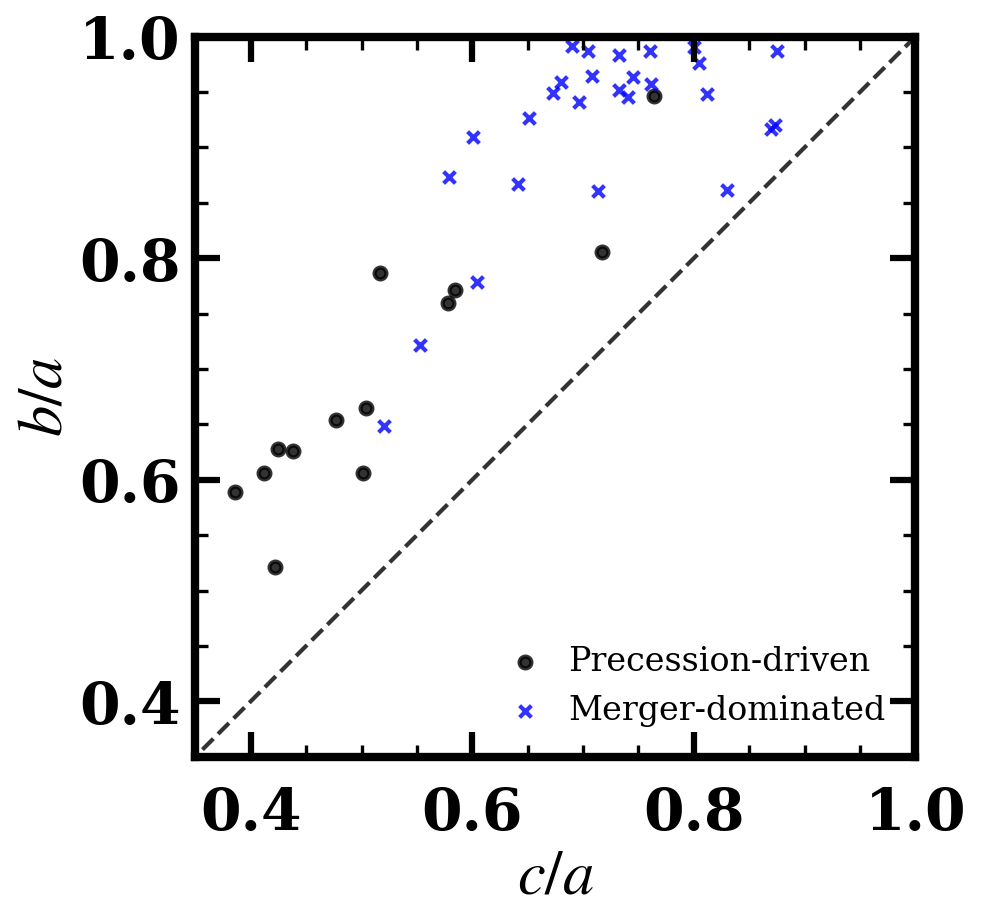}
   \caption{Relationship between the axis ratios $b/a$ and $c/a$ for dynamically hot galaxies at $z=0$. Black circles represent the dynamic hot galaxies in our sample, whose evolution is driven by strong precession, while the blue crosses denote the merger-dominated dynamic hot galaxies.} 
   \label{fig:axis_ratio}
\end{figure*}

To assess the impact of precession on galaxy morphology, we compare galaxies in our sample that evolve into dynamically hot systems ($f_{\rm hot}>0.8$) by $z=0$ (Figure~\ref{fig:WarpAngle}), with dynamically hot galaxies of comparable stellar mass whose evolution is predominantly driven by major mergers. These merger-dominated systems have experienced at least one major merger (defined as having a maximum merger mass ratio $\eta > 0.1$) during the period between the onset of their disk phase and $z=0$. 
We calculate the inertia tensor using all stellar particles within $2r_{e}$ to determine the principal axes of each galaxy. Figure~\ref{fig:axis_ratio} presents the ratio of the intermediate to major axis ($b/a$) as a function of the minor to major axis ratio ($c/a$). Despite the limited sample size, our galaxies exhibit systematically smaller values of $c/a$ and $b/a$ than merger-dominated systems. This suggests that the dynamic hot galaxies formed from disk galaxies undergoing strong precession preferentially evolve into prolate-like configurations. It is worth emphasizing that minor mergers and other dynamical processes may contribute to the structural transformations as well. These mechanisms can operate alongside, or even amplify, the disk precession itself. Therefore, the more prolate configurations seen here should be viewed as signatures of an evolutionary pathway that differs from that driven by violent major mergers. A more refined disentangling of the combined impacts of these processes is required and will be addressed in future work.

\section{Angular momentum alignment of cold gas streams for disk galaxies} \label{apd:angular_align}
We follow the evolution of the AM vectors of cold gas streams around the galaxy in our sample that are disk dominated ($f_{\rm hot}<0.5$) at $z=0$. Figure~\ref{fig:gas_track} of mian article presents two representative cases. Note that the other galaxies display a qualitatively similar behavior.
For each $z=0$ galaxy, we first select its cold gas cells ($\log_{10}(T/\mathrm{K}) < 5$) within $2r_{\rm e}$. We then trace back in time the evolution of the misalignment angle, $\alpha^{\rm gas}_{\rm star}$, between the AM vectors of these gas elements and the AM of the stellar component at $z=0$. The IDs of gas particles are not necessary because Vornoi cells continuously exchange mass and momentum across their interfaces, they do not correspond to persistence fluid elements. Therefore, we use Monte Carlo tracer particles \citep{Genel_tracer,Nelson_tracer} to follow gas evolution. As expected, at $z=0$ these gas elements lie in the rotational plane of the stellar disks, with the distribution of $\alpha^{\rm gas}_{\rm star}$ exhibiting a sharp peak  at values $<5^{\circ}$. At the time when these gas elements first cross into the inner $2r_{\rm e}$, the distribution of $\alpha^{\rm gas}_{\rm star}$ peaks around $5^{\circ}$, indicating that most inflowing gas streams have already settled into approximate co-rotation with the stellar disk. At larger distances of $\sim 50$ kpc from the galaxies, however, the distribution of $\alpha^{\rm gas}_{\rm star}$ is much broader. This implies that newly accreted cold gas initially arrives with a wide range of angular momentum orientations and is often strongly misaligned with the central stellar component. Once this gas penetrates into the inner region, its AM vectors are rapidly torqued and stabilized, becoming well aligned with the spin of the central galaxy.

\section{Theoretical analysis of the tidal torque and precession} \label{apd:theory}
\begin{figure*}
    \centering
    \includegraphics[width=8 cm]{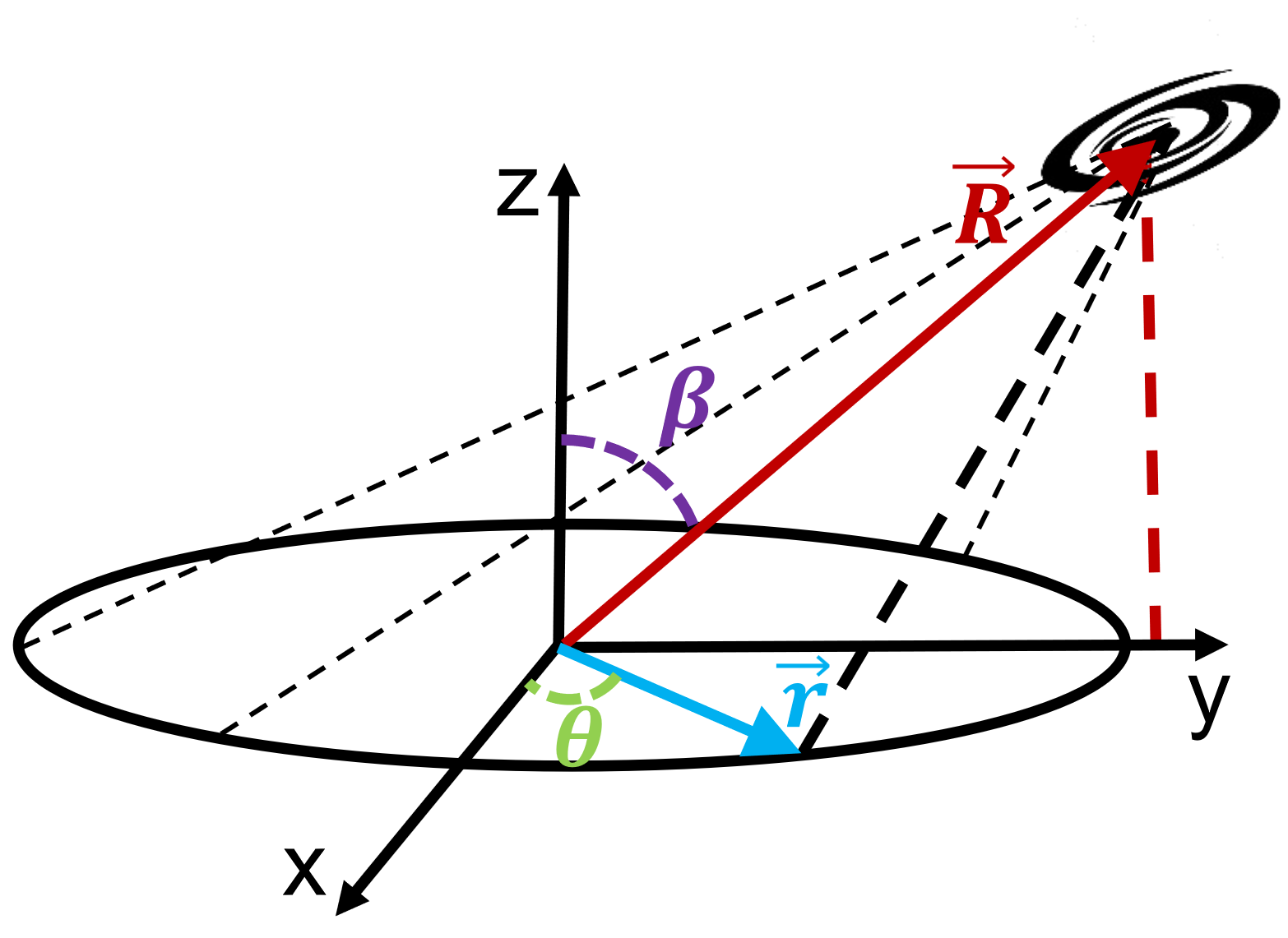}
    \caption{Schematic diagram of a coordinate system established with the $x\text{--}y$ plane in the disk plane, and the $z$-axis along the direction of central galaxy AM. A satellite galaxy is located in the $y\text{--}z$ plane, where $R$ denotes the distance from the central galaxy and $\beta$ is the polar angle relative to the $z$-axis. The disk annulus has a radius of $r$ and $\theta$ is the azimuthal angle of a mass element in the annulus.}
    \label{fig:coord}
\end{figure*}

In this section, we develop a simplified theoretical model to understand the precession induced by the tidal torque of local environment. We consider the simplest picture: a single satellite galaxy with a subhalo mass of $m_{\rm h,sat}$ close to a central disk galaxy. The satellite represents the anisotropic environment.
For simplicity, the satellite subhalo is approximated as a point-like particle and
the central galaxy disk is idealized as a series of concentric annuluses, each of which has a cylindrical radius $r$ and a stellar mass $m_{\rm *,cen}$. We set the AM vector of the central galaxy as the $z$-axis and put the satellite in the $y-z$ plane (Figure~\ref{fig:coord}).
The satellite is located at a distance of $R$ and has a polar angle of $\beta$ relative to the $z$-axis.

The total torque $\vec{M}$ on a disk annulus of radius $r$ induced by the satellite system at $\vec{R} = \left(0, R\sin{\beta}, R\cos{\beta} \right)$ can be obtained by integrating all moments of infinitesimal elements with a mass of $dm=m_{*,\rm cen}d\theta/2\pi$ at $\vec{r}(\theta) = (r\cos{\theta}, r\sin{\theta}, 0 )$ along the ring, where $\theta$ is the azimuthal angle:
\begin{equation}
    \begin{aligned}
        \vec{M} &= \int (\vec{r}\times \vec{g})dm,\\
        \vec{g} &= \frac{{\rm G} m_{\rm h,sat} }{{\left|\vec{R} - \vec{r}\right|}^3}(\vec{R} - \vec{r}).\\
    \end{aligned}
\end{equation}
Consequently, due to the symmetry of the ring, the integrals of the torque along the $y$-axis and the $z$-axis are necessarily zero. Therefore, the total torque consists of only the component along the $x$-axis:
\begin{equation}
    \begin{aligned}
        \vec{M} &= \int_{0}^{2\pi}  \frac{{\rm G}m_{\rm h,sat}}{{\left|\vec{R} - \vec{r}\right|}^3}\vec{r}\times \vec{R} \cdot \frac{m_{\rm *,cen} d \theta}{2\pi}\\
        &= \frac{{\rm G}m_{\rm *,cen}m_{\rm h,sat}}{2\pi} \int_{0}^{2\pi} \frac{rR\cos{\beta}\sin{\theta}d\theta}{(r^2+R^2-2rR\sin{\theta}\sin{\beta})^{3/2}}\hat{x},
    \label{eq:torque}
    \end{aligned}
\end{equation}
where $\hat{x}$ is the unit vector along the $x$ axis.
After applying the far-field approximation $r/R << 1$, Eq.~\ref{eq:torque} becomes:
\begin{equation}
    \begin{aligned}
        \vec{M} &= \frac{{\rm G}m_{*,\rm cen}m_{\rm h,sat}r}{2\pi R^{2}} \int_{0}^{2\pi} \frac{\cos{\beta}\sin{\theta}d\theta}{[1+(r/R)^2-2(r/R)\sin{\theta}\sin{\beta}]^{3/2}}\hat{x} \\
        &\approx \frac{{\rm G}m_{\rm *,cen}m_{\rm h,sat}r}{2\pi R^{2}}\hat{x} \int_{0}^{2\pi} \cos{\beta}\sin{\theta}[1+3(r/R)\sin{\theta}\sin{\beta}]d\theta  \\
        &= \frac{3{\rm G}m_{\rm *,cen}m_{\rm h,sat}r^2}{2 R^{3}}\hat{x}\sin{\beta}\cos{\beta}.
    \end{aligned}
    \label{eq:far_field_approx}
\end{equation}
Assuming a typical value of circular velocity $v_{\rm c}$ for the central galactic disk, the induced precession rate $\Omega_{\rm M}$, i.e., the angular velocity of the spin due to the torque, can be estimated as,
\begin{equation}
\begin{aligned}
    \Omega_{\rm M} &= \frac{M}{L} \approx \frac{\frac{3{\rm G}m_{\rm *,cen}m_{\rm h,sat}r^2}{2 R^{3}}\sin{\beta}\cos{\beta}}{m_{\rm *,cen} v_{\rm c}r}= \frac{3{\rm G}m_{\rm h,sat}r}{4 R^{3}v_{\rm c}}\sin{2\beta}. \\
    \end{aligned}
    \label{eq:OmegaM}
\end{equation}
Eventually, we derive Eq. \ref{eq:far_field_approx0}.

\begin{figure*}
   \centering
   \includegraphics[width=16cm]{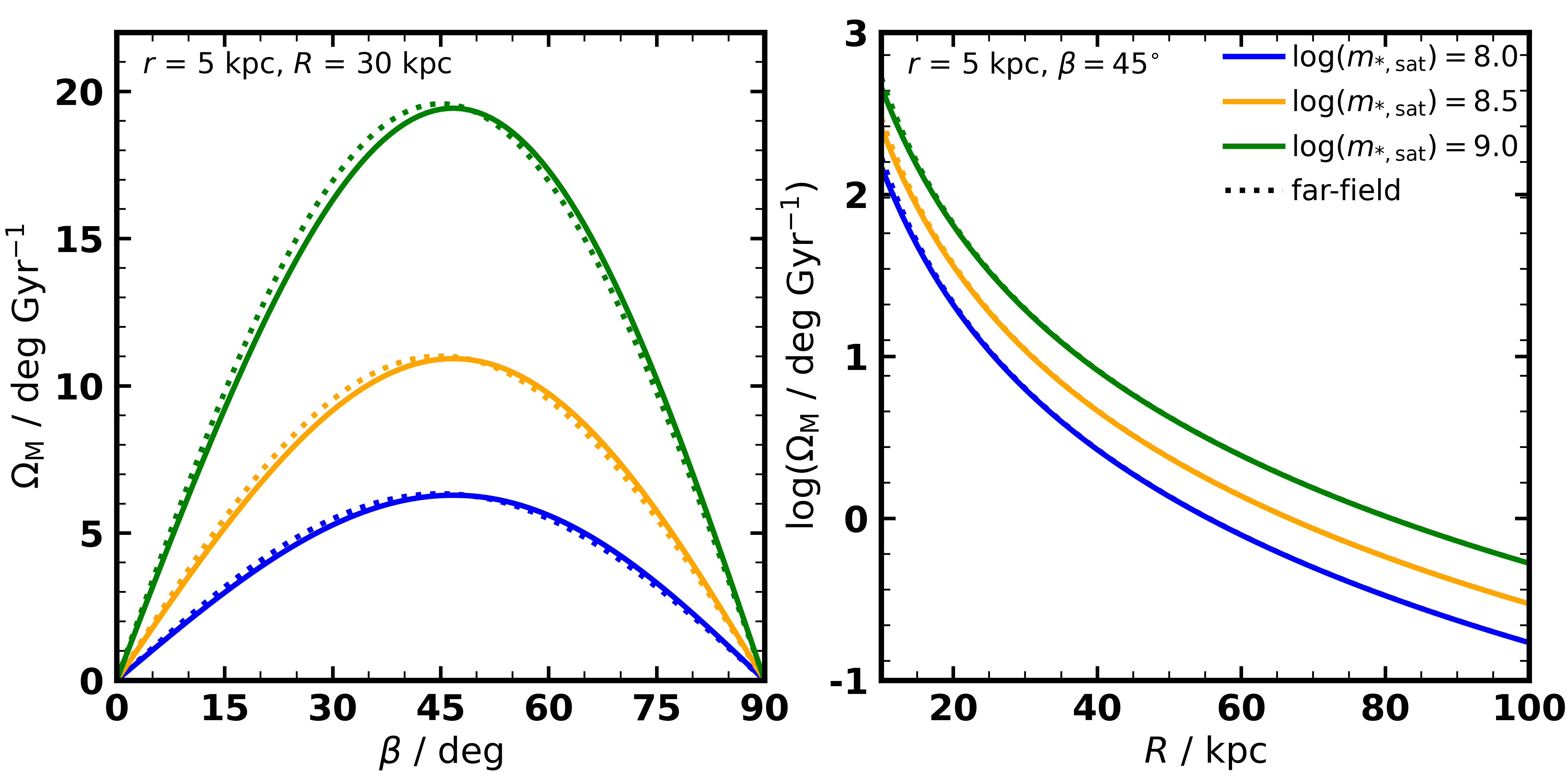}
\caption{Theoretical calculations for the precession of disk induced by the tidal torque from a satellite system using Eq.~\ref{eq:torque} (solid lines) and Eq.~\ref{eq:far_field_approx} (far-field approximation, dotted lines). 
The left panel shows $\Omega_{\rm M}$ as a function of $\beta$, while the right panel is for variable $R$, with the specified parameters displayed in the upper left corner.
The circular velocity of a selected disk ring at $r=5\ \rm kpc$ is set to be $v_{\rm c}=220 \ \rm km/s$.
Lines in different colors represent the results for typical host halos for varying satellite stellar masses as denoted in the legend. 
} 
    \label{fig:Omega_P_cen}
\end{figure*}

Figure~\ref{fig:Omega_P_cen} shows $\Omega_{\rm M}$ as a function of $\beta$ and $R$ for satellites with different stellar masses. The host halo masses ($m_{\rm h, sat}$) of these satellites are estimated using the stellar mass-halo mass relation given in the reference \cite{moster2010Constraints}. Here, we only consider a disk ring of $r=5$ kpc with a typical rotational velocity of $v_{\rm c} = 220 \ {\rm km}/{\rm s}$.
The solid and dotted lines are the results for the torques calculated using Eq.~\ref{eq:torque} and Eq.~\ref{eq:far_field_approx} (the far-field approximation), respectively. The small disparity between the solid and dotted lines proves that far-field is a reasonable approximation. $\Omega_{\rm M}$ drops rapidly with increasing $R$, as it is proportional to $R^{-3}$. 
The torque of asymmetric mass distribution can have a significant effect only when it is sufficiently close to the central galaxy.

The torque reaches its maximum around $\beta=45^{\circ}$, and vanishes when $\beta=0^{\circ}$ or $90^{\circ}$ or $180^{\circ}$.
A positive $\Omega_{\rm M}$ (for $\beta<90^{\circ}$) will result in an increase in $\beta$, while a negative $\Omega_{\rm M}$ (for $\beta > 90^{\circ}$, not shown in the figure) will lead to a decrease in $\beta$. 
Consequently, $\beta=0^{\circ}$ and $180^{\circ}$ are unstable equilibrium states and would move until the system reaches the stable point at $\beta=90^{\circ}$. 
Furthermore, Eq.~\ref{eq:OmegaM} can also be used to evaluate the precession of a satellite galaxy if we exchange the role of the central disk and the satellite system. This effect becomes more effective as the halo mass of the central galaxy is much larger compared to the satellite system, so that the disks of satellites preferentially point toward the direction of the central galaxy (as shown in the right panel of Figure \ref{fig:effects}. This effect is also important in regulating the AM of the accreted cold flow, as shown in Figure \ref{fig:gas_track}.

\bibliography{reference}{}
\bibliographystyle{aasjournalv7}



\end{document}